\documentclass[twocolumn,aps,prd,
preprintnumbers,
secnumarabic,amssymb,nofootinbib]{revtex4-1}
\pdfoutput=1

\usepackage{graphics}
\usepackage[dvips]{graphicx}
\usepackage{mathrsfs}
\usepackage{amssymb}
\usepackage{amsmath}
\usepackage{verbatim}
\usepackage{float}
\usepackage{slashed}
\usepackage{bbm}

\usepackage[english]{babel}
\usepackage{color,enumerate}

\newcommand{\tev}{\text{TeV}}
\newcommand{\FN}{\text{FN}}
\newcommand{\LFN}{\Lambda_{\rm FN}}
\def\ufn{U(1)_{\rm FN}}
\newcommand{\gev}{\text{GeV}}

\usepackage{graphicx,bm}
\def\nchi{\mathcal{Q}_\chi}

\newcommand{\lsim}{\lesssim}
\newcommand{\gsim}{\gtrsim}
\newcommand{\MET}{$E_T{\hspace{-0.47cm}/}\hspace{0.35cm}$}

\newcommand{\Lc}{\mathcal{L}}

\newcommand{\Mc}{\mathcal{M}}
\newcommand{\Oc}{\mathcal{O}}

\newcommand{\beq}{\begin{equation}}
\newcommand{\bea}{\begin{eqnarray}}
\newcommand{\eeq}{\end{equation}}
\newcommand{\eea}{\end{eqnarray}}
\newcommand{\bal}{\begin{align}}
\newcommand{\eal}{\end{align}}
\def\nn{\nonumber}
\def\half{\frac{1}{2}}
\newcommand{\la}{\langle}
\newcommand{\ra}{\rangle}

\newcommand{\rar}{\rightarrow}

\newcommand{\mdm}{m_\chi}
\newcommand{\mchi}{m_\chi}
\newcommand{\mrho}{m_\rho}
\newcommand{\msigma}{m_\sigma}
\def\ychi{y_\chi}
\def\lsh{\lambda_{sh}}
\def\ls{\lambda_s}

\usepackage{outlines}

\definecolor{darklightsabergreen}{rgb}{0.0, .49, 0.06}
\definecolor{orange}{rgb}{1.0, 0.5, 0.0}

\begin{document}

\title{Thermal dark matter via the flavon portal}

\author{Carlos Alvarado}
\email{calvara1@nd.edu}
\author{Fatemeh Elahi}
\email{elahif7@gmail.com}
\author{Nirmal Raj}
\email{nraj@nd.edu}
\affiliation{Department of Physics, University of Notre Dame, 225 Nieuwland Hall, Notre Dame, Indiana 46556, USA}

\begin{abstract}
{Dark matter (DM) is added to the Froggatt-Nielsen (FN) mechanism, and conditions for its successful freezeout identified. 
Requesting the FN scale $\LFN$ to be the cutoff of the theory 
renders freezeout scenarios surprisingly few. 
Fermionic DM is typically charged under $\ufn$, with the dominant annihilation channel a CP-even flavon + CP-odd flavon.  
A minimal case is when the DM-flavon coupling strength is $\Oc$(1), with several implications: 
(1) the DM mass is $\mathcal{O}$(100 GeV - 1 TeV), thanks to the WIMP coincidence,
(2) requiring perturbativity of couplings puts a lower {\em and} upper limit on the flavor scale, $2 \ {\rm TeV} \lsim \LFN \lsim 14~$TeV, on account of its relation to DM mass and couplings,
(3) DM is a ``secluded WIMP" effectively hidden from collider and direct detection searches.
Limits on the masses of dark matter and mediators from kaon mixing measurements constitute the best constraints, surpassing Xenon1T, Fermi-LAT, and the LHC. 
Future direct detection searches, and collider searches for missing energy plus a single jet/bottom/top, are promising avenues for discovery.
}
\end{abstract}

\maketitle

\section{Introduction}

The nature and origin of dark matter (DM) remain elusive. 
Since the Standard Model (SM) does not account for a DM candidate, it is natural to seek one in extensions of it devised to confront its other problems.
This approach enjoys an obvious merit: a single theory can account for (at least) two problems.
Thus, solutions to the electroweak (EW) hierarchy problem (e.g. weak scale supersymmetry and little Higgs) provide DM when a ``new physics" parity is imposed, right-handed neutrinos introduced to explain small neutrino masses, or axions introduced by the Peccei-Quinn resolution to the strong CP problem, may serve as DM -- and so on. 
Can DM be addressed in the problem of fermion flavors?

Fermion masses are hierarchical across many orders, and mix in peculiar patterns.
That these may be accidents of nature is an explanation we find unsatisfactory.
A simple alternative may be found in the mechanism of Froggatt and Nielsen (FN) \cite{Froggatt:1978nt}, that extends the SM gauge group with a (global or local) symmetry. 
The lightness of fermions $f$ is then arranged by mixing with heavy fermions $F$ vector-like under the new and SM symmetries: $ S F f + \LFN F \bar{F}$, where $S$ is the ``flavon", a scalar field acquiring a vev $v_{s}$ and breaking the symmetry.
The SM Yukawa matrices are now nothing but powers of $\epsilon \equiv \langle S\rangle/\LFN$ in an effective field theory (EFT), with $\epsilon$ usually fixed to the Cabibbo angle $\simeq 0.23$.
Both the mass and mixing hierarchies can be obtained now,
but flavor changing neutral currents (FCNCs) are inevitable.
To avoid constraints from FCNCs, it is found that $\LFN > 2$~TeV \cite{Bauer:2016rxs}.

It is to this picture that we wish to add DM. 
A profitable pursuit, one that gives experiments a well-motivated target,
is to identify the class of parameters that results in the correct relic abundance through $2 \rightarrow 2$ annihilations, in the spirit of such multi-parameter DM frameworks as supersymmetric neutralinos \cite{ArkaniHamed:2006mb,1412.4789,1510.03460}, minimal DM \cite{Cirelli:2005uq,0903.3381}, secluded WIMPs \cite{0711.4866}, effective WIMPs \cite{1307.8120,1402.7358}, and forbidden DM \cite{1505.07107,1608.05345}.
In other words, our first goal is to locate the ``relic surface".
Our other guiding principle is to add no more than a minimal set of mass scales to the FN mechanism. 
To begin with, we are not interested in the case of DM annihilations to the vector-like $F$'s, as this puts DM mass $> \LFN$ and generally out of current reach.
Thus, through operators suppressed by suitable powers of $\LFN$, DM must annihilate to SM fields and the flavon quanta that are obtained by expanding $S$ around its vev,
\begin{equation}
S = \frac{1}{\sqrt{2}}( v_s + \sigma + i \rho)~.
\label{eq:vevexpand}
\end{equation}

$\LFN$ is now the ``messenger scale" for DM interactions with SM and $S$, or in other words, the cutoff of our theory.
Following the FN procedure, we will arrange our EFT interactions by populating this scale with additional vector-like fermions.
One can broadly see where this leads if DM is a fermion singlet $\chi$. 
Assuming it to be odd under a $Z_2$ symmetry in order to avoid the operator $LH\chi$,
one may find that
interactions with all SM species must be suppressed by negative powers of $\LFN$, sometimes with extra suppression from factors of $v/\LFN$ (where $v$ is the Higgs vev) as well as from powers of $\epsilon$ (determined by the FN charges of SM and $\chi$). 
It can be verified -- and we will explicitly show it -- that these effects cause $\chi \overline{\chi} \rar$ SM SM to be too feeble, with the cross section $\la \sigma v \ra$ many orders smaller than  $(\eta) \ 2.2 \times 10^{-26} \ {\rm cm}^3 \ {\rm s}^{-1}$ (where $\eta=1 \ (2)$ for Majorana (Dirac) DM) that is required for the correct abundance.
DM interactions with the flavon, on the other hand, {\em need not} be $\LFN$-suppressed and may be arranged with marginal operators, as we shall show in this work. 
Couplings of $\mathcal{O}(0.1 - 1)$ are easily arranged,
rendering annihilations to the flavon particles $\sigma$ and $\rho$ a viable avenue.
Thus $\chi$ can be a ``secluded WIMP" \cite{0711.4866,Batell:2009vb}: it achieves the correct relic density by annihilating primarily to mediators (here $\sigma$ and $\rho$), while keeping direct couplings to SM small.
In this first paper, we will focus on a scenario where $\chi$ is charged under the flavor symmetry
and interacts with the flavon through a renormalizable Yukawa term $y_{\rm DM} \chi\chi S$;
the best constraints on this species of secluded WIMP come from indirect limits imposed by flavor experiments.
In a follow-up paper \cite{CAFENR}, we will extend our findings to cases where DM-flavon interactions are non-renormalizable, identify parametric families that lead to correct freezeout (while including SM annihilation channels that may become important), and 
derive all relevant constraints.

A most remarkable feature here is the hand played by the non-zero flavor charge of DM.
Due to this charge, DM mass must stem from symmetry-breaking as $\propto v_s \sim \epsilon \LFN$.
And parametrically, the cross section of $\chi$'s annihilation to the flavon mediators is given by $\la\sigma v\ra \sim y_{\rm DM}^4/\mdm^2$.
Since perturbative unitarity limits $\mchi$ from above \cite{Griest:1989wd}, an {\em upper} limit on $\LFN$ is imposed.
As a result, lower limits on $\LFN$ which may be placed by future flavor experiments or high-energy collider searches can potentially falsify our premise.
Moreover, since perturbativity at the flavor scale restricts $y_{\rm DM}$ to be $\Oc$(1), we know from the lore of the ``WIMP miracle" that, to obtain the characteristic $\la\sigma v\ra_{\rm th} = (\eta)~2.2 \times 10^{-26} \ {\rm cm}^3 \ {\rm s}^{-1}$, $\mdm$ must be $\lsim 10~$TeV.
Thus, although all the masses introduced here ($\LFN$, $v_s$, $\mdm$) were {\em a priori} free to be arbitrarily heavy, requiring correct freezeout puts 
them all within current experimental reach.
This attribute of a low-energy flavor-breaking scale emerging from a connection between DM and flavor was pointed out in \cite{Calibbi:2015sfa}.
It is comparable to \cite{Babu:1999me,0804.1753,1506.01719,1512.03458}, where a low $\LFN$ is obtained by breaking the flavor symmetry with electroweak Higgs doublets.
See also \cite{1204.1275} for model-independent constraints on low-scale flavor-breaking.

The above features will be spoiled if DM is a scalar, in which case it can have a renormalizable interaction with the Higgs doublet through a portal term: $|\chi|^2 |H|^2$.
Annihilations to the SM Higgs boson must dominate unless the coupling is tuned to be small, and there is no intimate relation between the DM abundance and the FN mechanism.
For these reasons our study will only focus on fermionic DM.

A Froggatt-Nielsen portal to DM was explored in \cite{Calibbi:2015sfa}, but the presence of the CP-odd flavon was omitted and emphasis was not placed on obtaining the correct relic abundance.
Here we will show that the CP-odd flavon plays a primary role in freeze-out.
The status and prospects of the CP-odd flavon were explored in comprehensive detail in \cite{Bauer:2016rxs}, whose results we will use extensively in this work.
For other works that explore the interface between flavor and dark matter, see \cite{1105.1781,Kamenik:2011vy,Dolan:2014ska,1408.3852,Kile:2013ola,Kamenik:2011nb,Varzielas:2015joa,Bishara:2015mha,Varzielas:2015sno,Galon:2016bka,1109.3516,1405.6709,1511.06293} and the references in \cite{1308.0584}.

This paper is laid out as follows.
We first review the FN mechanism in Sec.~\ref{sec:FN}.
DM is carefully incorporated into this set-up in Sec.~\ref{sec:Seeking}. 
We will find this a non-trivial task: we begin with a brief overview of simple models in Sec.~\ref{sec:DMcharge} and show them to be ineffective or unsatisfactory, before moving on to a successful model that requires $\chi$ to be charged under $\ufn$.
Sec.~\ref{sec:pheno} discusses constraints and future prospects, and Sec.~\ref{sec:concs} concludes the paper.

\section{Froggatt-Nielsen mechanism}
\label{sec:FN}

We begin with a brief review of the FN mechanism; for a more thorough review, see \cite{0910.2948}.
Ingredients relevant for embedding DM (performed in the next section) will be given emphasis.
The FN mechanism introduces an array of heavy vector-like fermions of mass $\LFN$ transforming under the SM gauge group as well as under a new symmetry that is either global, local or discrete;
we will use a global $\ufn$ for illustration.
The symmetry also transforms a new complex scalar $S$, the flavon, and 
all SM fermions excepting the top quark;
the Higgs doublet is neutral under it.
Conventionally, $S$ is assigned a $\ufn$ charge -1, which we assume hereafter.
The charge assignments ensure that in the theory below $\LFN$, fermions couple to the Higgs doublet only via non-renormalizable terms containing several powers of $S$ (or no power in the case of the top quark):

\beq
\mathcal{L}\supset y_{ij}^{(u)}\left(\frac{S}{\Lambda_{\FN}}\right)^{m_{ij}} \overline Q_i u_j \widetilde{H} + y_{ij}^{(d)}\left( \frac{S}{\Lambda_{\FN}}\right)^{n_{ij}} \overline{Q}_i d_j H~,
\label{eq:FN1}
\eeq 
where $\widetilde{H}=i\sigma_{2}H^{*}$ and the exponents $m_{ij}$, $n_{ij}$ are determined by the FN charges of the fermions.
For simplicity, we have assumed only quarks to be charged under the FN symmetry, though the mechanism can be easily extended to leptons as well. 
The $\ufn$ symmetry breaks if $S$ develops a vev $v_s$, giving rise to Yukawa couplings that are parameterically powers of $\epsilon = v_s /\sqrt{2}\Lambda_\FN$.
Thus, fermion masses and mixings originate in both electroweak symmetry breaking (EWSB) and flavor breaking, with their relative sizes set by the number of $\epsilon$ powers.
The size of $\epsilon$ is traditionally fixed by matching with measurements of the Cabibbo-Kobayashi-Maskawa (CKM) matrix: $\epsilon \simeq |V_{us}| \simeq |V_{cb}| \approx 0.23$.
Once the hierarchies are fixed to the right order this way, dimensionless $\mathcal{O}(1)$ coefficients $y_{ij}^{(u,d)}$ can bring the CKM entries and fermion masses to their measured values\footnote{The $y_{ij}$ must be complex to account for the CKM phase.}.

Thus far we have described the FN mechanism without explicit reference to the flavor group at work, which can be continuous, discrete, global, local, abelian, or non-abelian.
In our work we choose a global $\ufn$ for simplicity. 
Symmetry-breaking must introduce a potentially troublesome Goldstone boson $\rho$, disfavored by cosmological constraints if it couples to the SM \cite{1402.4108, 1309.5383,1509.06770}. 
This problem is evaded if the pseudoscalar $\rho$ acquires a non-zero mass through explicit breaking\footnote{Alternatively, $\ufn$ may be either 
(i) gauged, which may however introduce anomalies as the left- and right-handed fermions are charged differently, or
(ii) discretized, in which case there is no Goldstone boson. 
Through the $Z_N$-preserving operator $S^{N}/\Lambda_{\text{FN}}^{N-4}$, where $N$ is the dimension of the $Z_N$ group, one has $m_{\rho}^{2} \sim \epsilon^{N-4}v_{s}^{2}$. 
Successful FN models require $N \leq 16$ since the up quark demands an $\epsilon^{8}$ suppression, implying a very light $\rho$ for $v_s \sim \Oc$(TeV) that is already excluded by flavor constraints \cite{Bauer:2016rxs}.} in the potential:
\begin{align}
V(H,S) =& - \mu_s^2 |S|^2 + \lambda_s |S|^4 \nonumber\\
&+ \lambda_{sh} |S|^2H^{\dag}H - b^2 ( S^2 + \text{H.c.})~,
\label{eq:potential} 
\end{align}
giving rise to the physical masses post-minimization
\begin{equation} \msigma^2 = 2 \lambda_s v_s^2, \ \ \ \ \ \  m_\rho^2 = 4 b^2~,
\label{eq:sigmarhorelations}
\end{equation}
with $b^{2}>0$ and $v \simeq$ 246 GeV.
Though $b^2$ is a free parameter, as it is the only term explicitly breaking $\ufn$, it is multiplicatively renormalized and can be naturally smaller than the other scales here. 
Thus we require $\mrho$ to lie below $\LFN$ and assume the mass hierarchy in Ref.~\cite{Bauer:2016rxs}:
\begin{equation*} 
\mrho < \msigma \simeq v_s < \LFN~.
\label{eq:flavonhierarchy}
\end{equation*}

Eq.~\ref{eq:FN1} determines the Yukawa couplings $(g_s)_{ij}$ of $\sigma$ and $\rho$ with quark pairs of families $i$ and $j$, written out explicitly in Appendix \ref{sec:allYukawas}.
These couplings generate tree-level FCNC processes, due to which the FN set-up confronts limits from measurements of meson mixing, meson decays, and top quark decays, with the strongest constraints imposed by the neutral kaon mixing CP-violation parameter $\epsilon_K$ \cite{Bauer:2016rxs}.
The latter constrains the masses of $\sigma$ and $\rho$, which can be translated to limits in \{$\ls,~v_s,~\mrho$\} space\footnote{Through the rest of the paper we take into account the fact that the right-hand sides of Eq.~\ref{eq:sigmarhorelations} are 2 $\times$ those in Ref.~\cite{Bauer:2016rxs}}.
The constraint is weak at $\mrho \simeq \msigma \approx 200$~GeV due to an accidental cancellation in the Wilson coefficients of $\Delta F = 2$ operators, but when $\mrho \gg 200$~GeV,
the contribution of the flavon quanta goes as $(g_\sigma)^2_{sd}/\msigma^2 \propto (\ls v_s^4)^{-1}$ (from Eqs.~\ref{eq:sigmarhorelations} and \ref{eq:gschichicoupsexplicit}).
Thus a lower limit on $v_s$ and $\LFN = v_s/(\sqrt{2}\epsilon)$ may be obtained if we require the coupling $\ls$ to be perturbative. 
From Ref.~\cite{Bauer:2016rxs}, we have 
\bea
\nn \ls \leq 4\pi \ \ \ &\Rightarrow& \ \ \  v_s \geq 670~{\rm GeV}~\\
&\Rightarrow& \LFN  \geq 2.07~{\rm TeV}~,
\label{eq:vslimit} 
\eea
which will serve for the purposes of this paper as absolute lower limits on $v_s$ and $\LFN$.

Once we embed DM into the FN picture,
the above constraints may also restrict DM masses.
Therefore, we will revisit these constraints in more detail in Sec.~\ref{sec:pheno},
while in the next section we proceed to our principal task of adding DM.
 
\section{Incorporating DM}
\label{sec:Seeking}

\subsection{General model-building}
\label{sec:DMcharge}

Having described the interactions of the flavon quanta $s = \sigma, \rho$ with SM states, we turn to our central program of incorporating fermionic DM into the FN setup.

The following are general considerations to keep in mind before we delve into the details of model-building.
\begin{itemize}
\item
As mentioned in the Introduction, we impose a $Z_2$ symmetry under which SM fields and $S$ are even and $\chi$ is odd.
This prevents operators of the form $(S/\Lambda)^k H L \chi$ that could result in DM decay, where $k \geq 0$ is an integer.

\item
We also mentioned in the Introduction that annihilations to SM species 
are suppressed by inverse powers of the cutoff scale and $\epsilon$, 
and that successful freezeout is only obtained through annihilations to the flavons.
Hence our emphasis in the following will be on DM interactions with flavons.
All these low-energy interactions are assumed to arise from vector-like fermions integrated out at the scale $\LFN$.
Some of these vector-like fermions, $F_\chi, \bar{F}_\chi$, must be charged odd under the $Z_2$ symmetry that stabilizes DM. 
In principle these vector-like fermions could have arbitrary masses, but we have chosen for them a common mass $\LFN$ in line with our objective of keeping the number of new mass scales at a minimum.
Thus the theory at high scales would appear as
\begin{equation*}
\Lc \supset  S \chi \bar F_\chi +  \LFN F_\chi  \bar F_\chi~.
\end{equation*}

\item
Annihilations of a DM pair into $\sigma + \rho$ are $s$-wave, whereas annihilations into $ \sigma \sigma$  and $\rho \rho$ are $p$-wave.
We see this from parity considerations.
The fermion-pair initial state has $P = (-1)^{L+1}$, thus the
$L = 0$ transition is allowed for the parity-odd
$\sigma \rho$ final state, and forbidden for the parity-even $\sigma\sigma$ and $\rho\rho$ final states.

\item
As $\chi$ can in principle be either neutral or carry an (arbitrary) $\ufn$ charge, we must seek a successful model by sifting through the possibilities.
\end{itemize}

With these considerations, we now explore the freeze-out of DM neutral or charged under $\ufn$.

\subsubsection{$\ufn$-neutral DM}
\label{subsubsec:ufnneutral}

Here DM has a bare Majorana mass $\mchi$,
and connects to the flavon via the lowest dimension operator $\chi\chi|S|^{2}/\LFN$.
Assuming CP-violating phases vanish, the interactions 
$$  \chi\chi \rho^2/\LFN~, \ \  \chi\chi \sigma^2/\LFN~, \ \ \epsilon \ \chi\chi \sigma~ $$
are obtained from Eq.~\ref{eq:vevexpand}.
These give rise to the annihilation of $\chi$ to pairs of $\sigma$ and $\rho$, which are $p$-wave-suppressed.
In addition, annihilation through the first two is $\LFN$-suppressed, 
and that through the third is non-trivial to arrange:
$\chi$ must be heavier than $\sigma$ to kinematically allow it, and at least an order of magnitude lighter than $\LFN$ for the EFT to be valid.
From Eq.~\ref{eq:sigmarhorelations}, this means $v_s < \mdm \ll \LFN$ for the quartic coupling $\lambda_s \sim 1$. 
However, this is not possible since $v_s \simeq \epsilon \LFN$. 
Of course, the special hierarchy $\msigma < \mdm \ll \LFN$ may be contrived if $\lambda_s \ll 1$, but we do not pursue this possibility since we expect the region of viability to be small for light $\sigma$ in the face of flavor constraints. 
Also, as we mentioned in the Introduction, we wish to keep the introduction of new mass scales minimal (violated in this case by the introduction of $\mdm$).

To summarize, $\ufn$-neutral DM can possibly lead to successful freeze-out through $p$-wave annihilations to $\sigma$ pairs, but only a small region would survive flavor constraints. 
A larger region of viability is possible if DM can annihilate to $\sigma + \rho$ through the $s$-wave instead.
In the next sub-section we will show that DM charged under $\ufn$ is more successful in this respect.

\subsubsection{$\ufn$-charged DM}
\label{subsubsec:ufncharged}

DM charged under $\ufn$ must acquire its mass, and interactions with the flavon quanta $s$, via symmetry-breaking.
Specifically, DM acquires a Dirac mass and couplings to $S$ through the operator
\beq
y_\chi\left(\frac{S}{\LFN}\right)^n S \chi_a \chi_b~,
\label{eq:masterop}
\eeq
given by
\begin{equation}
\mdm =  \frac{\ychi}{\sqrt{2}} v_s \epsilon^n, \ \ g_{s \chi \chi} = (n+1)\frac{\ychi}{\sqrt{2}} 
\epsilon^n~,
\label{eq:genparams}
\end{equation}
where $n$ determines $\nchi \equiv$ the collective charge of $\chi_a$ and $\chi_b$ as
\beq
\nn \nchi = (n+1)/2~.
\label{eq:nnchirelation}
\eeq
Without loss of generality, we take $\ychi$ to be real.
The only free parameters in this set-up are now 
\bea
\nn & {\rm Scales} &:~~ \LFN,~b^2, \\
\nn & {\rm Charge} &:~~ \nchi,\\
& {\rm Couplings~~}&:~~ \ychi,~\lambda_s,~\lambda_{sh}~.
\label{eq:freeparams}
\eea

It is among these parameters that we must find successful freezeout conditions and identify the relic surface.
For our phenomenological treatment in Sec.~\ref{sec:pheno}, we neglect  
$\lambda_{sh}$, for it plays little role in our freezeout: as we will show in Sec.~\ref{subsec:U1DMcharged}, its influence by means of turning on a small Higgs-$\sigma$ mixing is negligible.
 
\begin{figure}[t]
\centering 
\includegraphics[scale=0.52]{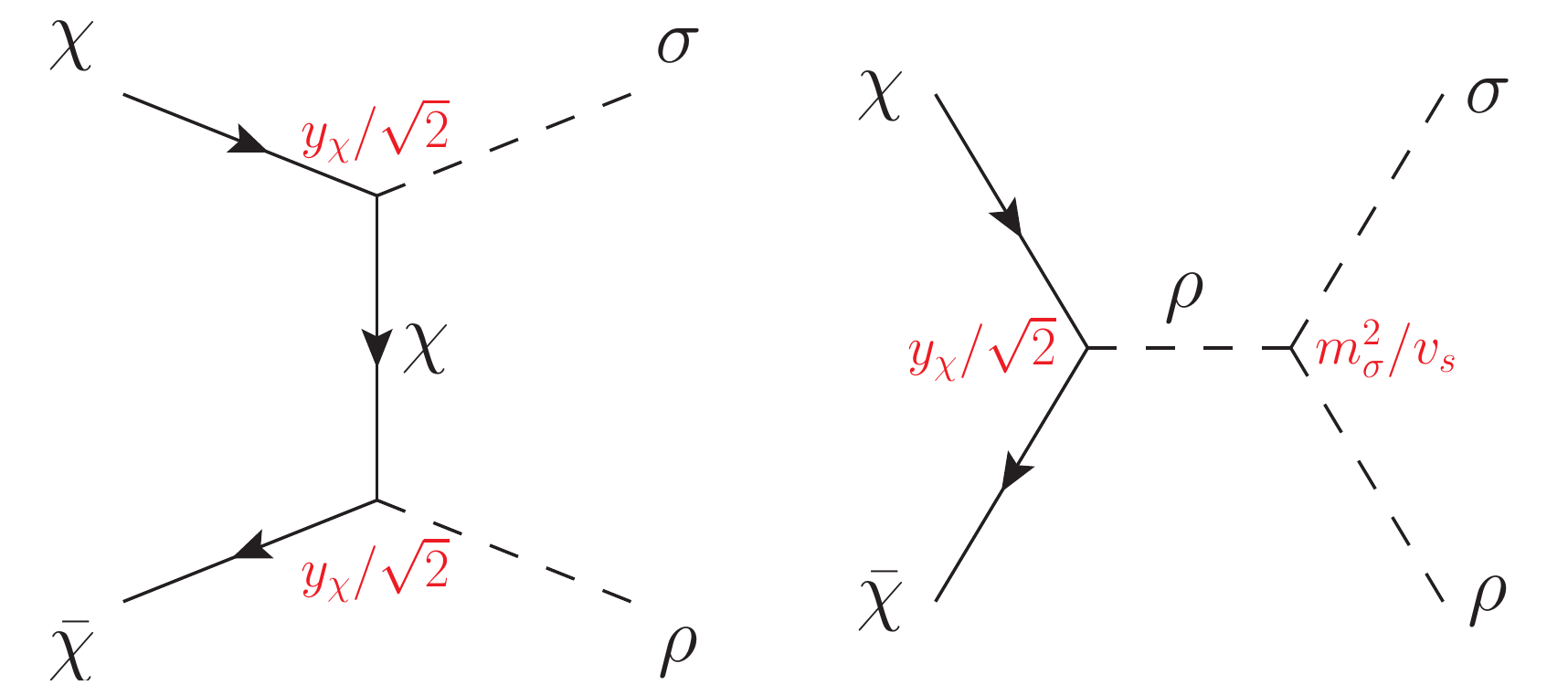} 
\caption{Feynman diagrams contributing to the $s$-wave annihilation of DM to a CP-even + CP-odd flavon.
When kinematically allowed, this channel dominates for DM charged $1/2$ under $\ufn$.
See text for more details.
}
\label{fig:feyndiags}
\end{figure}

We now proceed to find our desired conditions.
First, we notice that Eq.~\ref{eq:genparams} allows for the $s$-wave process 
$\chi \bar{\chi} \rightarrow \sigma \rho$.
Both the $s$- and $t$-channel diagrams in Fig.~\ref{fig:feyndiags} contribute, and lead to the annihilation cross section given in Appendix~\ref{app:DMformulae}. 
For $\mrho \ll \mdm$, 
this is schematically 
\bea
\nn \la \sigma v \ra \sim \frac{1}{s}|\Mc|^2 &\sim& \frac{1}{4\mdm^2} f'' \\
 &\sim& \frac{\epsilon^2}{4\LFN^2} f',
\label{eq:genannXS}
\eea
where $f''$ and $f'$ are functions of $\ychi, \ls, n$ and $\epsilon$. 
The above equation implies our set-up can give us a potential {\em upper} limit on  the Froggatt-Nielsen scale $\LFN$ when we require the thermal cross section $\la 
\sigma v \ra_{\rm th} = 4.4 \times 10^{-26} \ {\rm cm}^3 \ {\rm s}^{-1}$. 
 This must happen when we require that the coefficient $\ychi$ be perturbative 
 ($\ychi \leq 4\pi$).
 In the following we will derive this upper limit on $\LFN$ for a few select cases.

Let us begin our investigation of DM annihilations with the case of $n=0 \ (\Rightarrow \nchi = 1/2$).
Here 
\begin{eqnarray}
\nn & & \mchi = \frac{\ychi}{\sqrt{2}}~v_s, \ \ g_{s \chi \chi} = \frac{\ychi}{\sqrt{2}} \\
 &\Rightarrow& \la \sigma v \ra \simeq \frac{3}{2048\pi}\frac{\ychi^2}{v^2_s}~,
\label{eq:nzeroparams}
\end{eqnarray}
where in the second line we have used Eq.~\ref{eq:sWave} and set $\mdm = \msigma$ for simplicity. 
This can certainly lead to successful freezeout, provided $v_s$ is not so large as to make $\ychi$ non-perturbative. 
This is not a cause for concern, since Eq.~\ref{eq:vslimit} implies $\ychi \geq 1.9$ if we require the correct abundance at $\mdm \gg \mrho$.
(For $\mdm \sim \mrho$, there is no lower bound on $\ychi$.)
Annihilations to $\sigma + \rho$ are kinematically allowed so long as $\msigma + \mrho < 2 \mdm \Rightarrow  2 b^2 < (\ychi - \sqrt{\ls})^2 v_s^2$.
Requiring $\ychi \leq 4\pi$ gives $\LFN \leq 13.65$~TeV, which is allowed by the limit in Eq.~\ref{eq:vslimit}.

To sum up, we have found our first successful freezeout scenario without contriving a compressed mass spectrum.
Our work will chiefly concern this scenario, for reasons that will become apparent when we inspect the effect of increasing $n$.

As we increase $n$, Eq.~\ref{eq:genparams} implies that $\chi$ gets lighter, reducing the phase space available for annihilation to $\sigma + \rho$.
(One may try to recover some phase space by tuning $\ls$ small and making $\sigma$ light, but at the cost of tension with kaon mixing constraints.)
Thus the $p$-wave flavon modes ($\sigma \sigma$ and $\rho \rho$) and SM modes gain in importance.
Moreover, inserting Eq.~\ref{eq:genparams} into Eq.~\ref{eq:genannXS}, $\la \sigma v \ra \propto \ychi^2 (n+1)^4 \epsilon^{2n}/\LFN^2$ in the $\ls \rightarrow 0$ limit, implying that as we increase $n$, the upper bound on $\LFN$ from $\ychi$ perturbativity gets stronger.
Eventually this upper bound will fall below the lower bound in Eq.~\ref{eq:vslimit}.
For $\mdm = \msigma$, this occurs at $n = 4$, which gives us an important condition for successful freezeout:
\beq
\nn n \leq 3, \ {\rm or} \ \nchi \leq 2.
\eeq
Conditions on the parameters in Eq.~\ref{eq:freeparams} that render desired annihilation modes kinematically allowed may be derived in a straightforward manner from Eqs.~\ref{eq:sigmarhorelations} and \ref{eq:genparams}.

Our next task is to show that, after imposing these conditions and locating our relic surface, our set-up is quite viable in the face of dark matter experiments.
To this end, we pick a single scenario for phenomenological study, $\nchi = 1/2 \ (n = 0)$.
Our choice is motivated by the following reasons.

(1) As we just showed, the $n = 0$ case provides the maximum phase space for the channel $\chi \bar{\chi} \rightarrow \sigma \rho$, allowing it to dominate the annihilation over a large parametric region.
This simplifies the phenomenological analysis. 

(2) As Eq.~\ref{eq:masterop} is a marginal operator for $n=0$, we may relax the assumption that $Z_2$-odd vector-like fermions with a common mass $\LFN$ generate DM-flavon interactions at low energies, and assume no more than the presence of a pair of  dark fermions with a combined $\ufn$ charge of unity\footnote{We also assume that neither individual charge $\nchi^a, \nchi^b$ is zero. 
If one of the $\chi_i$ is $\ufn$-neutral, a Majorana mass $\half M_m \chi_i\chi_i$ and the operator $|S|^2|\chi_i\chi_i/\LFN$ are allowed, confounding the freezeout analysis and potentially introducing physical complex phases.}.

\begin{figure}[tb!]
\centering 
\includegraphics[width=.4\textwidth]{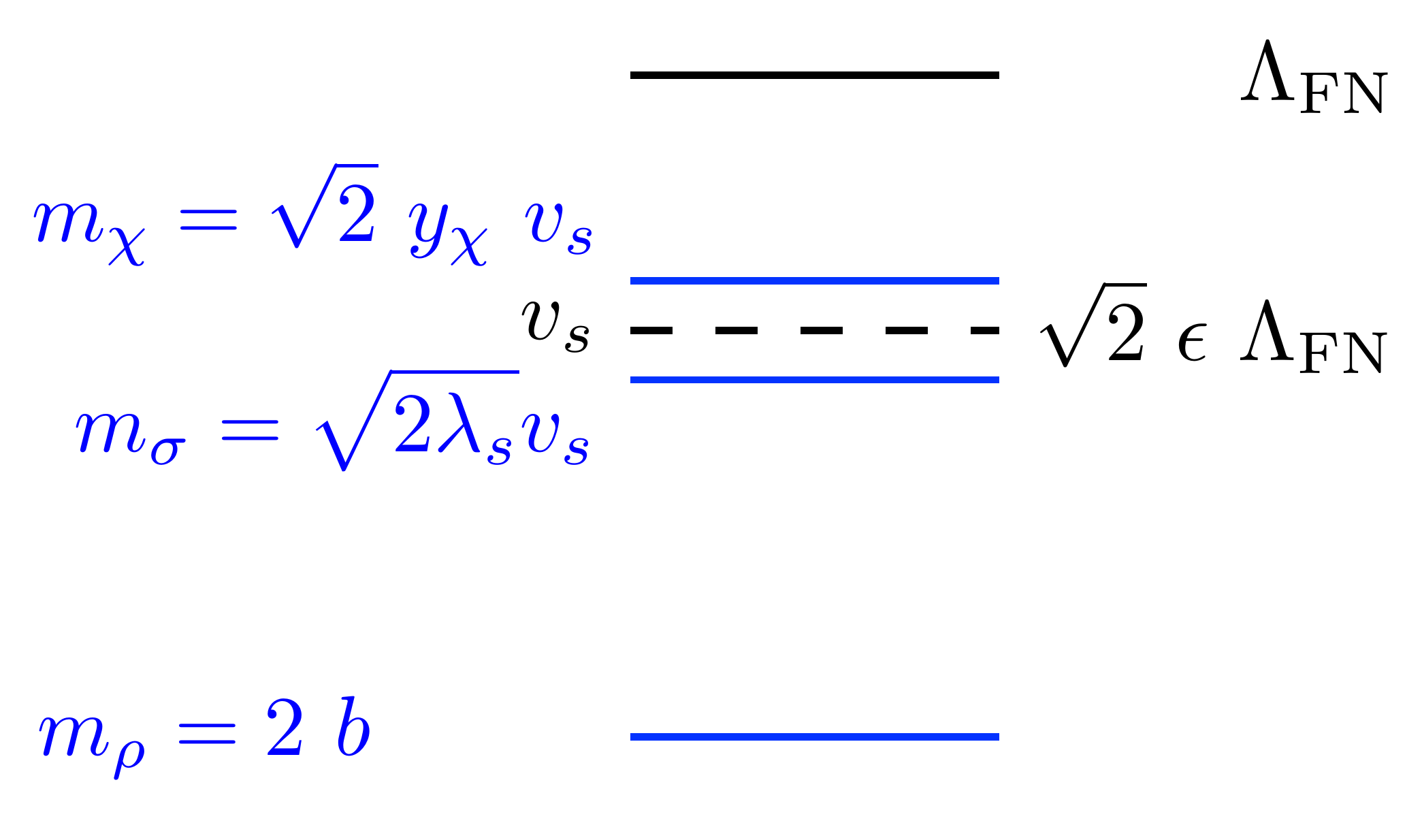}\\
\caption{The spectrum studied in this work.
The $\ufn$ symmetry breaks at $\sqrt{2} \epsilon$ below the Froggatt-Nielsen scale $\LFN$, giving masses to dark matter $\chi$ and the CP-even flavon $\sigma$ at the symmetry-breaking scale $v_s$.
Shown for illustration is a hierarchy in which $\mdm > \msigma$.
The mass of the CP-odd flavon $\rho$, acquired through a freely tunable explicit symmetry breaking parameter $b^2$, is assumed $< 2 \mdm - \msigma$ in order to allow for DM annihilations to $\rho + \sigma$.
}
\label{fig:spectrum}
\end{figure}

The spectrum of scales in our scenario is sketched in Fig.~\ref{fig:spectrum}.
In general $\mdm, v_s$ and $\msigma$ reside at a common $\Oc({\rm TeV})$ scale,  while $\mrho$, a free parameter, can be much lower.
The relation between these masses and scales will play a decisive role in our phenomenology.

\subsection{Flavon mode domination: an illustration}
\label{subsec:U1DMcharged}

\begin{figure*}
\centering 
\includegraphics[width=0.45\textwidth]{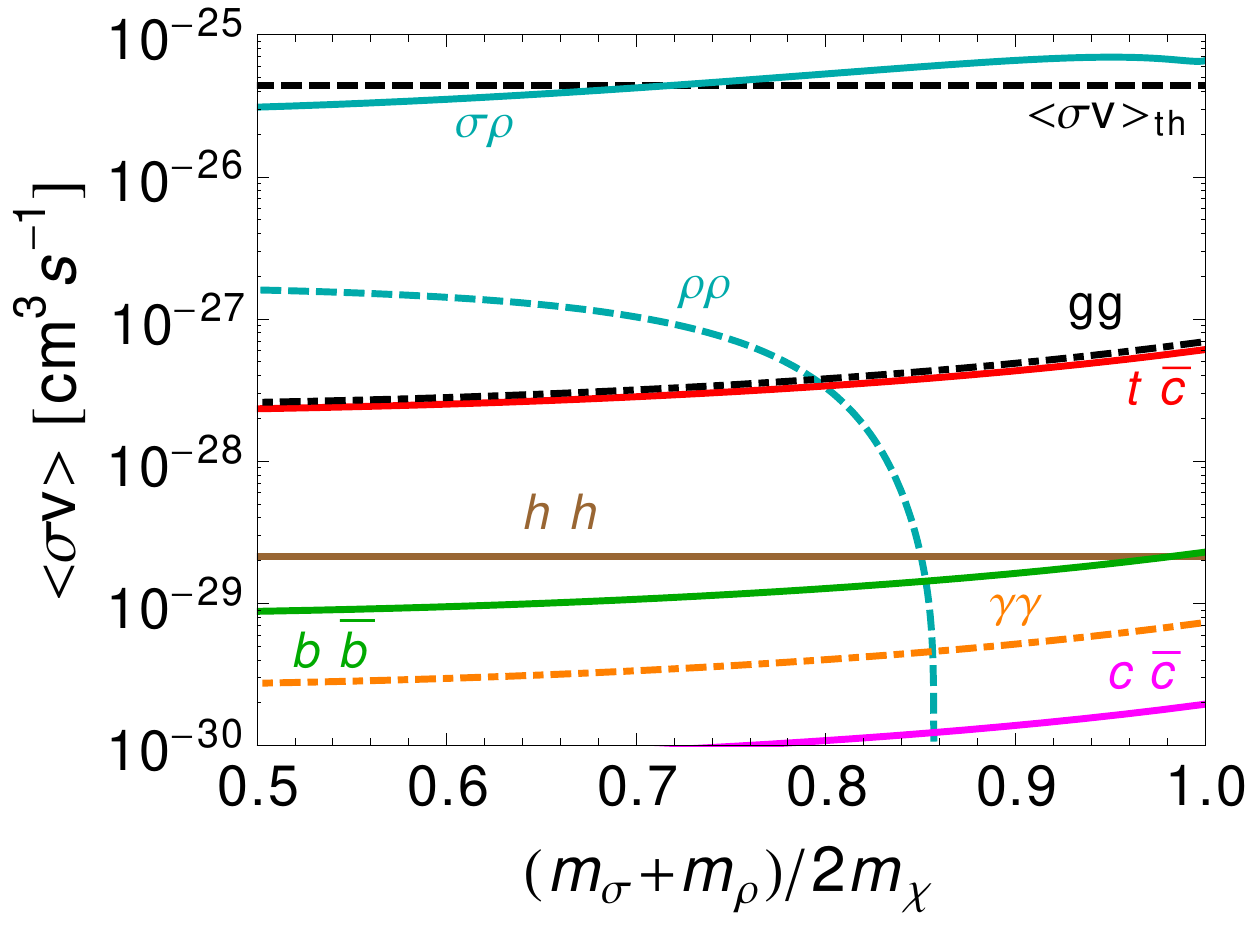}
\quad \quad
\includegraphics[width=0.45\textwidth]{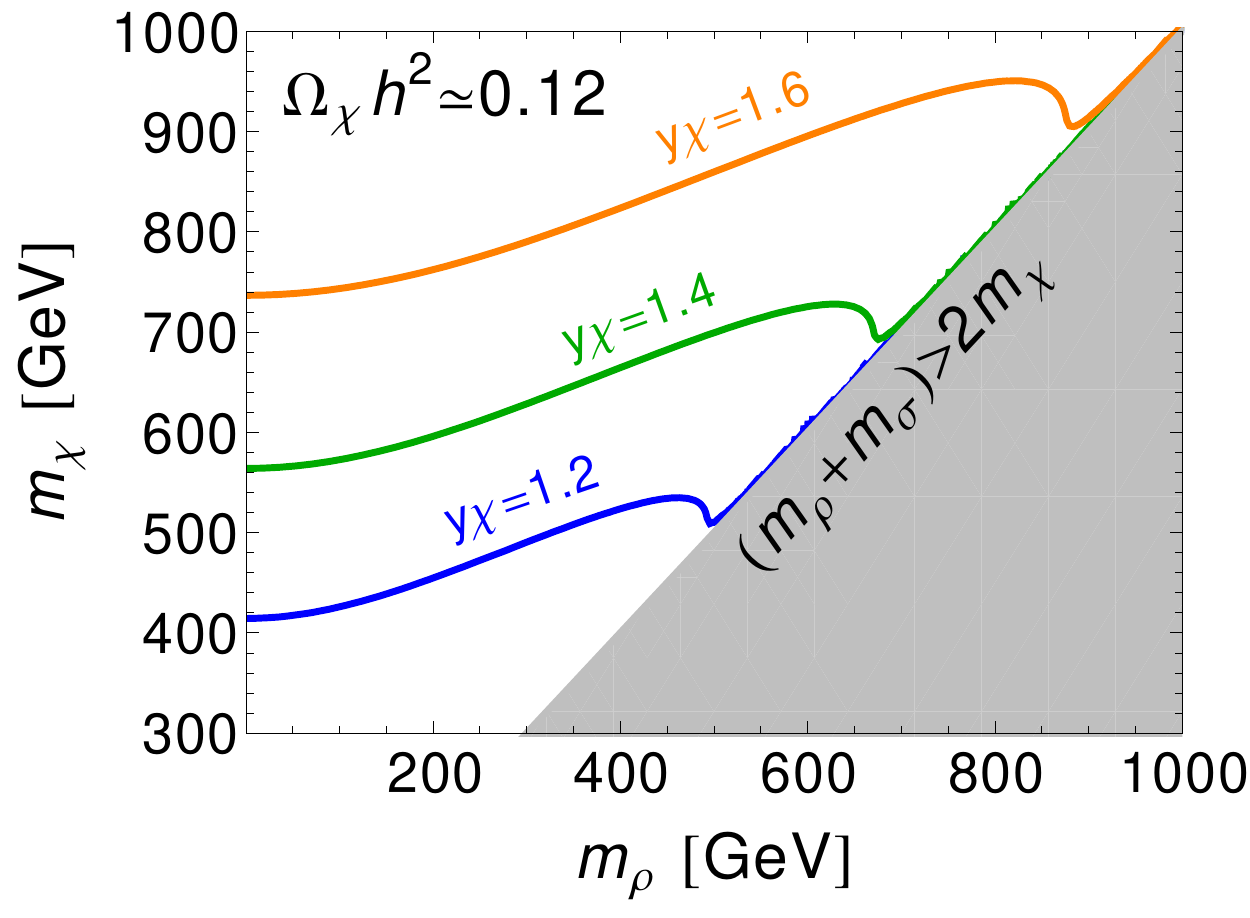}
\caption{\textbf{Left:} Cross sections of various DM annihilation channels as a function of $(\msigma+\mrho)/2\mchi$, keeping $\mdm = 950~\gev, \ \ychi = 1.4, \ \lambda_s = 0.25$, and $\lambda_{sh} = 0.1$.
Annihilations to the CP-even $\sigma$ + CP-odd $\rho$ are seen to dominate over all other modes. 
\textbf{Right:} Contours of $\ychi$ resulting in the correct relic abundance, fixing $\msigma = \mdm$. 
See text for more details.
}
\label{fig:ann}
\end{figure*}

In the Introduction we had estimated that DM annihilations to all SM final states will be suppressed.
We had also surmised that freezeout will be dictated by $s$-wave annihilations to flavon quanta.
In the previous sub-section, after identifying $\ufn$-charged DM as a workable scenario, we derived freezeout conditions ignoring SM modes and including only flavon modes.  
We now demonstrate the accuracy of our assumptions by quantifying these estimates, which form the crux of our paper. 

The left panel of Fig.~\ref{fig:ann} shows the $\la \sigma v \ra$ of various annihilation modes against the ratio $(\mrho + \msigma)/2\mdm$, with the thermal cross section $\la\sigma v\ra_{\rm th} = 4.4 \times 10^{-26} \ {\rm cm}^3 \ {\rm s}^{-1}$ shown for reference.
We have chosen $\mdm=950~\gev, \ \ychi = 1.4, \ \lambda_s = 0.25$ and $\lambda_{sh} = 0.1$ for illustration; this puts $\msigma$ at $678~\gev$.
The relevant SM modes, $h h, \ t\bar{c}, \ b \bar{b}, \ c \bar{c}, \ g g$, and $\gamma \gamma$ are plotted in brown, red, green, magenta, dot-dashed black, and dot-dashed orange respectively, the flavon modes $\sigma \rho$ and $\rho \rho$ in solid blue and dashed blue.
Our parametric range kinematically forbids the $\sigma \sigma$ mode, but allowing it does not change our conclusions.
We checked our calculations against {\tt MicrOmegas} 4.3 \cite{MuOmega} and found very good agreement.

Let us begin our task by first inspecting the SM final states.
Annihilations to Higgs bosons ($\chi \overline{\chi} \rightarrow \sigma^* \rightarrow h h$)  proceed through the $\lsh$ vertex in Eq.~\ref{eq:potential}, and suppressed by the twofold effect of its $p$-wave nature and the large mass of $\sigma$ in the propagator.
This is why the cross section is three orders of magnitude below $\la\sigma v\ra_{\rm th}$.
Even for $\lsh$ as large as 1, the above effects would keep the cross section
at a factor of 100 below the $\la\sigma v\ra_{\rm th}$.
In general, turning on the coupling $\lsh$ would induce $h$-$\sigma$ mixing,
introducing potential constraints from LHC Higgs measurements.
However, due to the hierarchy between the mass scales $m_h \sim v$ and $\msigma \sim v_s$,
the mixing angle comes out to be $\lsim 0.1$, which is safe from these constraints. 
For this reason, and because $\lsh$ plays no role in the freezeout, we consistently neglect it throughout the rest of the paper.
As a consequence, we will also not be concerned with ($p$-wave suppressed) annihilations to the electroweak bosons that would have been prompted by a non-zero $\lambda_{sh}$.

Annihilations to SM fermions are highly suppressed as well.
These must proceed through flavon mediation in the $s$-channel;
since both $\rho$ and $\sigma$ couple to fermion pairs through the Higgs doublet (as seen in Eq.~\ref{eq:FN1}), a factor of $(v/\LFN)^2 < 10^{-2}$ appears in the cross section.
The relative contributions of the fermion modes is determined by the number of $\epsilon$ powers in the DM-flavon coupling, which is shown in Appendix~\ref{sec:allYukawas}.
Re-writing Eq.~\ref{eq:YukawasFlavon} (up to $\Oc(1)$ coefficients) as

\begin{align}
\nn g_{s}^u = \frac{1}{v_s}
\begin{pmatrix} 
    8 m_u              &  \epsilon m_c  &  \epsilon^3 m_t \\ 
     \epsilon^3 m_c & 4 m_c              &  \epsilon^2 m_t \\
     \epsilon^5 m_t &  \epsilon^2 m_t   & 0 \,
     \;
     \end{pmatrix}~,
     \\
     g_{s}^d = \frac{1}{v_s}
\begin{pmatrix} 
    7 m_d            &  \epsilon m_s &  \epsilon^3 m_b \\ 
     \epsilon m_s & 5 m_s              &  \epsilon^2 m_b \\
     \epsilon m_b &  \epsilon^2 m_b   & 3 m_b
\end{pmatrix}~,
\label{eq:gschichicoupsexplicit}
\end{align}
we see that except for $t\bar{c}, \ b\bar{b}$ and $c\bar{c}$, 
all other fermion modes are too feeble.

The presence of a global $\ufn$ anomaly in our set-up gives rise to the annihilation modes $\chi\bar{\chi}\rightarrow\rho^{*}\rightarrow gg, \gamma \gamma$. 
Calculating the $\rho g g$  and $\rho \gamma \gamma$ couplings using the color and electromagnetic anomaly coefficients that originate from quark triangle diagrams \cite{1106.2162},
we find the $gg$ cross section comparable to $t \bar{c}$, and the $\gamma \gamma$ cross section 100 times smaller. 

We are now left with annihilations to two flavons.
The $\sigma \rho$ mode, contributing $> 95\%$ to the total cross section, is dominantly $s$-wave (with the $p$-wave contribution so negligible as to vary the solid blue curve only minutely).
As advertised in Sec.~\ref{subsubsec:ufncharged}, this annihilation can proceed through an $\Oc(1)$-sized coupling to produce $\la \sigma v \ra_{\rm th} = 4.4 \times 10^{-26} \ {\rm cm}^3 \ {\rm s}^{-1}$.
We see this clearly in the solid blue curve.
The $\rho \rho$ (and $\sigma\sigma$) mode is $p$-wave.
Consequently, its cross section is suppressed by about an order of magnitude with respect to the $\sigma \rho$ mode.
The cross section also drops sharply as $\mrho$ approaches $2\mdm - \msigma$ and shrinks the phase space open for annihilation.

Finally, the right panel of Fig.~\ref{fig:ann} shows, in the $\mrho$--$\mdm$ plane, contours of $y_\chi$ that result in successful freezeout.
We have set $\msigma = \mdm$ in this plot, which from Eqs.~\ref{eq:sigmarhorelations} and \ref{eq:nzeroparams} implies $\ls = \ychi^2/4$ along each contour.
As $\mrho$ is raised, the phase space available for $\chi \bar{\chi} \rightarrow \sigma \rho$ is reduced, requiring a slight increase in $\mdm = \msigma$ to recover the thermal cross section.
One also observes that larger couplings are needed for heavier DM to overcome the $\mdm^{-2}$ suppression of the annihilation cross section.

In the next section we explore the various signals and constraints of our set-up, and show that our parameter space on the relic surface is by and large allowed by flavor and dark matter experiments.
Searches best suited for finding our scenario are also identified and discussed.

\section{Phenomenology}
\label{sec:pheno}

Since our DM gets its relic abundance effectively by annihilating to mediators, it is a ``secluded WIMP" \cite{0711.4866}, generally hidden from the Standard Model and hence expected to be probed poorly by direct detection and colliders.
And though our annihilation is $s$-wave, allowing our set-up to submit to indirect detection searches, our DM is generally too heavy to produce sufficient photonic flux  to be seen.
However, flavor-changing processes can competently probe the mediators $\rho$ and $\sigma$.

In this section we will discuss the constraints on our scenario from these various experiments, and predict our future prospects.
We will begin with flavor experiments, recasting the findings of Ref.~\cite{Bauer:2016rxs} in our parameters and finding bounds on $\mdm$ and $\mrho$.
The most stringent limits here are from kaon mixing measurements.
Next we discuss constraints from direct detection, and show that future searches would reach regions that are allowed by kaon mixing.
Then we briefly discuss the poor (current and future) sensitivity of indirect detection. 
Finally, we show that current LHC limits too are weak, and explore promising DM signatures for Run 2.

\subsection{Meson mixing}
\label{sec:kaon}

\begin{figure*}
\begin{center}
\includegraphics[width=0.45\textwidth]{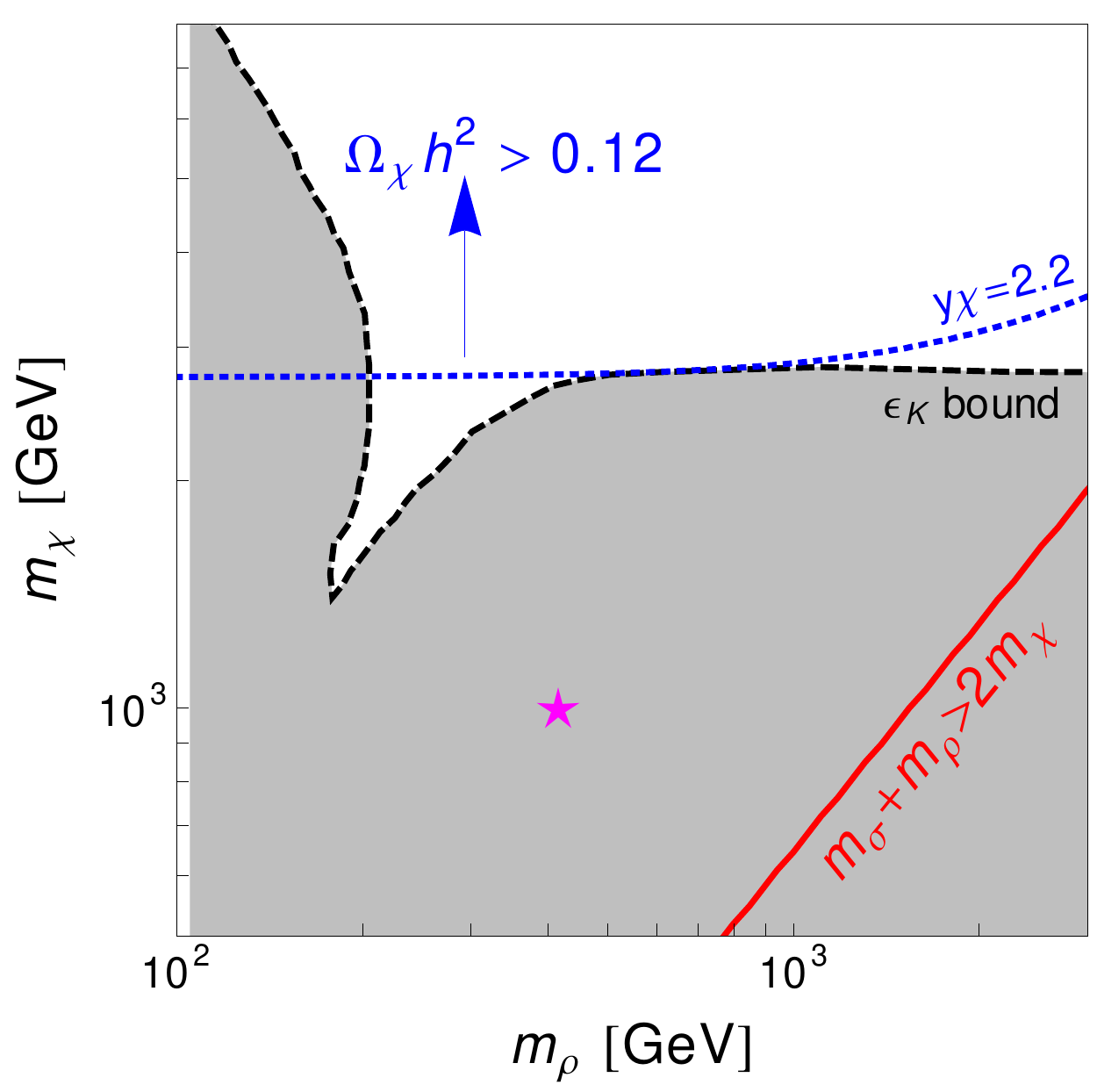}
\quad \quad
\includegraphics[width=0.45\textwidth]{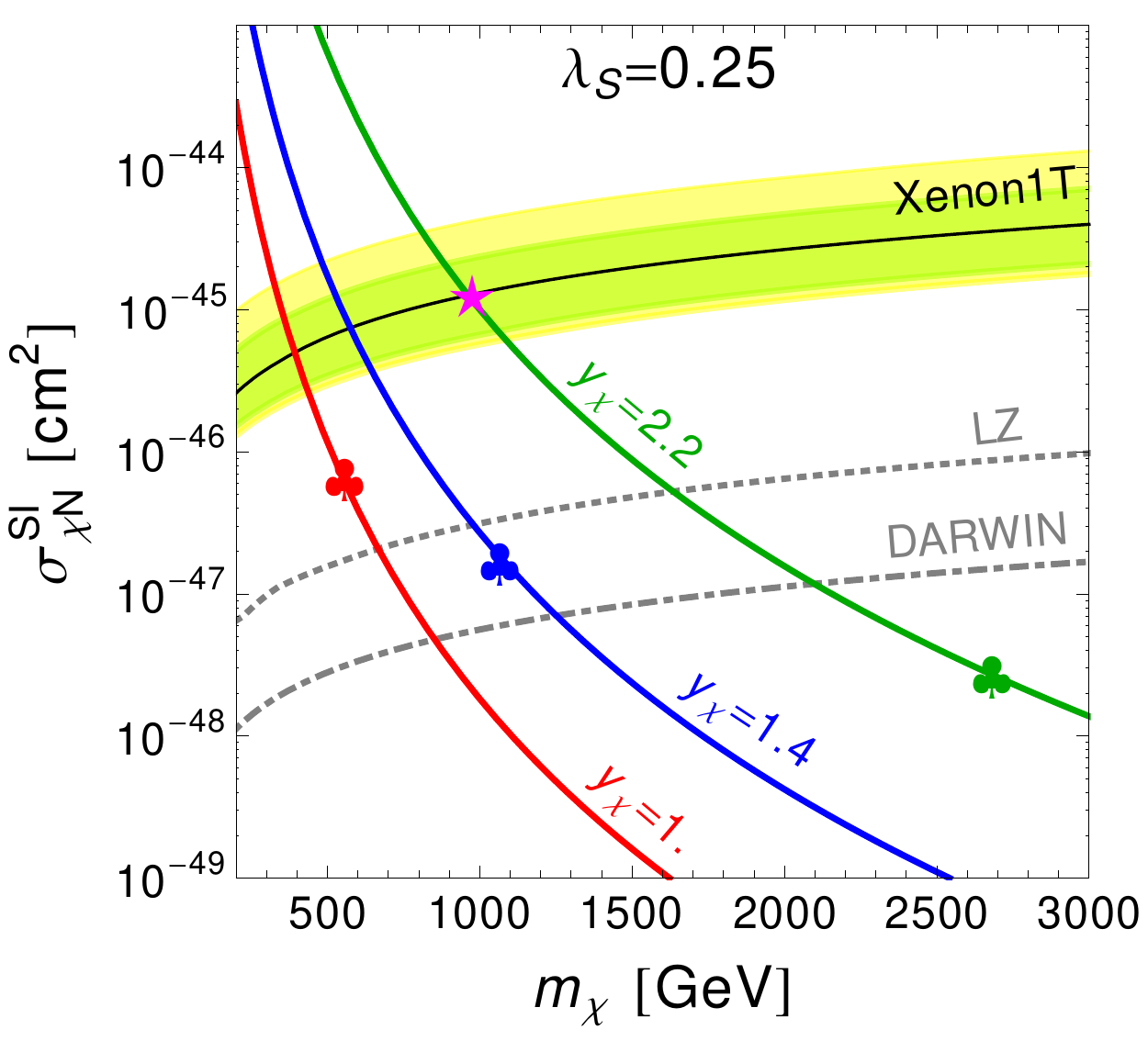}
\caption{   \textbf{Left:} 95\% C.L. bounds from $\epsilon_{K}$ measurements (dark shaded region excluded) in the $\mdm$--$\mrho$ plane, at $\lambda_{s}=0.25$ and $\ychi = 2.2$.
The blue dashed curve is a contour of $\Omega_\chi h^2 = 0.12$; the region above it leads to overabundant DM. 
The region to the right of the red line is kinematically forbidden. {\bf Right:} Spin-independent DM-nucleon scattering cross sections as a function of $m_{\chi}$ for $y_{\chi}$ = 1.0, 1.4 and 2.2 (red, green, and blue curves), versus current constraints from Xenon1T (solid black) and future sensitivities at LZ and DARWIN (dotted and dot-dashed). 
The cloverleaf on each $y_{\chi}$ curve shows the $m_{\chi}$ that leads to $\Omega_\chi h^2 = 0.12$.
\textbf{Both panels:} The pink star in the right-hand panel at $\mdm =$ 960 GeV is replicated at the same mass in the left-hand panel -- this is as an example indication that regions excluded by direct detection are usually even more deeply excluded by kaon mixing constraints.}
\label{fig:limits}
\end{center}
\end{figure*}

Both the CP-even $\sigma$ and CP-odd $\rho$ exhibit flavor violating couplings at tree level (Eq.~\ref{eq:gschichicoupsexplicit}), generating FCNCs
and incurring low-energy constraints from meson mixing in neutral $K, B$ and $D$ systems \cite{Bauer:2016rxs}.
The strongest limits come from kaon mixing
since the SM contribution is rendered small by the GIM mechanism, i.e. it is both loop- and CKM-suppressed.
The CP-violation parameter $\epsilon_K$ is dominated by short-distance contributions that can be accurately calculated, whereas the observable $\Delta m_K$  suffers from a large theoretical uncertainty due to unknown long-distance contributions.
For this reason $\epsilon_K$ is generally expected to provide the best constraints.
Ref.~\cite{Bauer:2016rxs} recast results from the \textbf{UT}\textit{fit} collaboration \cite{Bona:2007vi} onto the $\mrho$--$v_s$ plane, and showed that this is indeed true.
Their choice of $\Oc(1)$ coefficients and $\Oc(1)$ phases that appear in the Yukawa textures of Eq.~\ref{eq:HiggsYukawas} does play a role in this result,
but one must, for the reasons explained above, expect $\epsilon_{K}$ to outconstrain the CP-preserving $\Delta m_K$ for most Yukawa textures\footnote{We thank F. Bishara (M. Bauer) for raising (clarifying) this point.}.
We use this result to show our constraints in the $\mrho$--$\mdm$ plane in the left panel of Fig.~\ref{fig:limits}, taking advantage of the relation between $\mdm, \ychi$ and $v_s$ in Eq.~\ref{eq:nzeroparams}.
However, the reader must keep in mind that the limits discussed in the following, as well as the limit quoted in Eq.~\ref{eq:vslimit}, may be weakened for some choices of the $\Oc(1)$ coefficients and phases in the Yukawa couplings.

We fix $\ychi = 2.2$ in Fig.~\ref{fig:limits} and show with a dashed curve a contour of $\Omega_\chi h^2 = 0.12$.
In the regions above this contour, DM is overabundant for this $\ychi$.
The effect of varying this coupling is seen in Fig.~\ref{fig:ann}, however, it must be remembered that raising or lowering $\ychi$ would correspondingly tighten or loosen the $\epsilon_K$ bound on $\mdm$.

The dark shaded region is excluded at 95\% C.L. by the $\epsilon_K$ measurement, with an illustrative $\ls$ value of 0.25.
As explained in Sec.~\ref{sec:FN}, the bound comes from tree-level contributions to $\Delta F = 2$ operator Wilson coefficients, which depend on $\mrho$ and $\msigma$.
For $\mrho > 200~$GeV, these contributions scale as $\ls^{-1} v_s^{-4} \propto \ls^{-1} (\ychi/ \mdm)^{4}$, hence giving a flat bound across $\mrho$. 
 It will prove useful to recast this as a scaling of the lower bound on $\mdm$ in 
 terms of the couplings:
\beq
\mdm|_{\epsilon_K} \propto \frac{\ychi}{\ls^{1/4}}~.
\label{eq:Kmixscaling}
\eeq
For $\mrho < 200~{\rm GeV} \ll \msigma$, the flavon contributions to the Wilson coefficients scale as $(v_s\mrho)^{-2} \propto (\mdm\mrho/\ychi)^{-2}$, giving a bound on $\mrho$ that falls inversely with $\mdm$.
The dip feature seen between these two regions comes from an accidental cancellation in the Wilson coefficient at $\msigma = \mrho$ due to destructive interference.

This plot illustrates clearly that regions favored by our freezeout scenario are quite viable vis-\`a-vis flavor constraints.
The choice of 2.2 is the smallest $\ychi$ that allows our scenario to escape the $\epsilon_K$ bound for all $\mrho > 200~$GeV.
We find that for $1.2 \leq \ychi \leq 2.2$, our set-up is viable when the relic contour is trapped in the dip feature.
For $\ychi < 1.2$, we are completely excluded.

Since regions where our relic contours are excluded mostly correspond to $\mrho > 
200~$GeV, where the scaling in Eq.~\ref{eq:Kmixscaling} roughly applies,
for our discussions below we will use this equation for making comparisons with 
constraints from DM experiments.
In the following sub-sections we will pay particular attention to whether these experiments have future sensitivities to parametric regions not excluded by flavor probes.

\begin{figure}
\begin{center}
\includegraphics[width=0.45\textwidth]{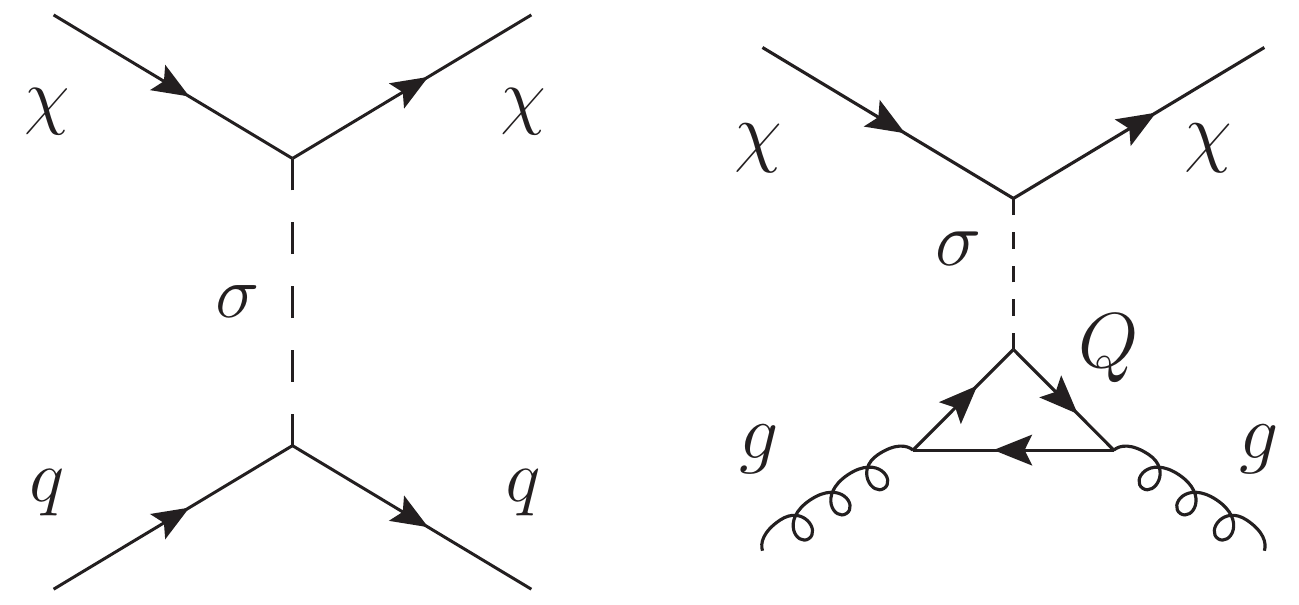}
\caption{Diagrams contributing to spin-independent scattering with nucleons. 
Here $q = u, d, s$ and $Q = c, b, t$.}
\label{fig:feynDD}
\end{center}
\end{figure}

\subsection{Direct detection}
\label{subsec:DD}

DM can scatter with nucleons through flavon exchange, potentially introducing constraints from direct detection searches.
It is well-known that fermion DM scattering with nucleons via pseudoscalar mediator exchange produces a spin-dependent cross section that is velocity-suppressed \cite{1305.1611}.
Thus, only the exchange of the CP-even $\sigma$ is relevant to our scenario.
As sketched in Fig.~\ref{fig:feynDD}, this leads to scattering via light quark operators at the tree level and via gluon operators through heavy quark loops.
In general, we expect the rates to be small since
they will be suppressed by a large $m_{\sigma}$ ($\sim v_s$).

Setting aside considerations of relic density for the moment, we now inspect the constraints.
We computed our spin-independent cross section $\sigma_{\rm SI}$ using the formulae
in Appendix~\ref{app:DMformulae}.
From Eqs.~\ref{eq:genparams}, \ref{eq:gschichicoupsexplicit} and \ref{eq:XSSI}, this cross section scales as 
\beq
\sigma_{\rm SI} \propto \mu^2_{\chi N} m_p^2 \left(\frac{\ychi^8}{\ls^2\mdm^6}\right)~.
\label{eq:XSSIscaling}
\eeq

The right panel of Fig.~\ref{fig:limits} plots $\sigma_{\rm SI}$ against $\mdm$ for three choices of $y_{\chi}$: 1.0 (red), 1.4 (blue) and 2.2 (green), and fixing $\lambda_s$ to 0.25.
The $\ychi^8$ scaling may be seen by comparing among these curves at some $\mdm$.
The 90\% C.L  exclusion cross sections (with their 1 and 2 $\sigma$ bands) set by Xenon1T \cite{Aprile:2017iyp} are provided for reference.
Due to the scaling in Eq.~\ref{eq:XSSIscaling}, our limit tightens with $\ychi$.
In general, we expect direct detection to constrain the DM mass poorer than kaon mixing.
For instance, one may read off the plot that for $\ychi = 2.2$, $\mdm \gsim 960~$GeV. This is a weaker bound than the kaon mixing one, illustrated by the pink star at the corresponding $\mdm$ in both the left and right panels of Fig.~\ref{fig:limits}.
Variations in $\ychi$ will not change this behavior.
Since the exclusion cross section rises only gently across $\mdm$ in this region, 
we expect from Eq.~\ref{eq:XSSIscaling} that the limit on $\mchi$ scales as $\ychi^{4/3}$.
On the other hand, from Eq.~\ref{eq:Kmixscaling}, we know that the kaon mixing $\mdm$ bound scales as $\ychi$, which is not much slower than the direct detection scaling.

Our future direct detection prospects are quite interesting.
To check these, we compare our $\sigma_{\rm SI}$ with the projected reaches of the LUX-ZEPLIN (LZ) \cite{1509.02910} and DARWIN \cite{1606.07001} experiments, provided in the figure.
 For $\ychi = \{1.0, 1.4, 2.2\}$, LZ is sensitive to $\mdm \lsim \{0.7, 1.0, 1.6 \}$~TeV and DARWIN to $\mdm \lsim \{0.8, 1.2, 2.1 \}$~TeV.
Amusingly, there emerge three distinctive future prospects of our relic surface for the three $\ychi$ choices.
We show this by placing a cloverleaf on our $\sigma_{\rm SI}$ curves for each $\ychi$ at the $\mdm$ that gives $\Omega_\chi h^2 = 0.12$ 
(the part of the $\sigma_{\rm SI}$ curve to the left/right of the cloverleaf corresponds to DM freezing out under/over-abundantly. 
Also, as seen in the left panel of Fig.~\ref{fig:limits}, the relic contour is near-insensitive to $\mrho$, and mostly picked by $\mdm$). 
By scanning the cloverleaves (at $\mdm \simeq \{0.55, 1.1, 2.7\}$~TeV) and the points where our $\sigma_{\rm SI}$ curves intersect with LZ and DARWIN, we conclude that

(a) both LZ and DARWIN can reach $\ychi = 1.0$,

(b) LZ cannot, but DARWIN can, reach $\ychi = 1.4$, 

(c) neither LZ nor DARWIN can reach $\ychi = 2.2$.

As we had mentioned in the previous sub-section, for $\ychi \in [1.2, 2.2]$ our relic contour evades the $\epsilon_K$ bound in the dip feature.
Thus we have a small range of $\ychi$ that gives the correct abundance, is currently viable with respect to all constraints, and is discoverable by future direct detection searches.
(Note that although $\ychi > 2.2$ evades the $\epsilon_K$ bound even without help from the dip feature, it is not discoverable at direct detection.)
Though these results were obtained after fixing $\ls$, varying it would not change the broad conclusions.

We end this sub-section on a final note.
Throughout the above, we have set $\lambda_{sh} = 0$ following the motivation in Sec.~\ref{subsec:U1DMcharged}, but had we turned it on and enabled mixing with the Higgs boson, some contribution to scattering cross sections due to Higgs and gauge boson exchange may have arisen; however, these are completely negligible due to the small mixing angles quoted in Sec.~\ref{subsec:U1DMcharged}.

\subsection{Indirect detection}
\label{subsec:ID}

Fermi-LAT observations of gamma rays from stacked dwarf galaxies \cite{1310.0828} have set 90\% C.L. limits on DM annihilation cross sections that can reach $\la \sigma v \ra_{\rm th} = (\eta)~2.2 \times 10^{-26} \ {\rm cm}^3 \ {\rm s}^{-1}$, potentially affecting our set-up since our annihilation $\chi \overline{\chi}  \rightarrow  \sigma \rho$ is mainly $s$-wave.
However, these limits are generally much weaker than the $\epsilon_K$ bound discussed in Sec.~\ref{sec:kaon}.
Consider the strongest Fermi-LAT bound, that from a 100\%~$b\bar{b}$ final state: for $\la \sigma v \ra = \la \sigma v \ra_{\rm th}$, DM mass $\gsim 100~$GeV.
Although our annihilation scenario is different -- SM final states are products of $\sigma$ and $\rho$ decay, i.e. we have cascaded, as opposed to direct, annihilations -- the corresponding limit on $\mdm$ must not be too far from 100 GeV.
Indeed, we find that $\mdm \gsim 175$~GeV from a naive recasting that assumes (a) the integrated photon flux from our DM cascaded annihilation equals that from direct annihilation to $b\bar{b}$, (b) equal masses, and hence decay branching ratios, for $\sigma$ and $\rho$.
This limit falls well short of the one corresponding to the smallest $\mdm$ spared by the $\epsilon_K$ bound: $\mdm \gsim 760$~GeV, which occurs at $\ychi = 1.2$ as explained in Secs.~\ref{sec:kaon} and \ref{subsec:DD}.
Future indirect detection prospects too are dim.
By recasting the sensitivities provided in \cite{1310.2695}, we find that only $\mdm \lsim 325~$GeV is within Fermi-LAT's reach.

These conclusions, based on order-of-magnitude estimates, would hold against variations in $\ls$. 
But they would change dramatically if we choose $n > 0~(\nchi > 1/2)$ in Eq.~\ref{eq:masterop}, since Eq.~\ref{eq:genparams} implies $\mdm$ can now be lighter and hence in the range of Fermi-LAT.
This warrants a througher investigation of the indirect detection phenomenology of our $n > 0$ parametric families, which we undertake in \cite{CAFENR}.
 
\subsection{LHC}
\label{subsec:LHC}

The $13~\tev$ LHC can potentially probe our scenario.
The collider propects of $\rho$ have already been thoroughly explored in Ref.~\cite{Bauer:2016rxs}.
To summarize these briefly:
\begin{itemize}
 
\item the collider phenomenology primarily arises from the fact that $\rho$ couples strongest to bottom-bottom and top-charm; this can be seen in Eq.~(\ref{eq:gschichicoupsexplicit}). 

\item the production rates are dominated by some combination of $b\bar{b} \rightarrow \rho, \ gb \rightarrow b\rho$ and $gc\rightarrow t \rho$, depending on $\mrho$,

\item the decay branching ratios of $\rho \rightarrow b \bar{b}$ and $\rho \rightarrow t \bar{c}$ dominate,

\item at a 100 TeV collider, studies of top $\rightarrow \rho \ +$~charm can exclude the region $\mrho \leq 175$~GeV, whereas $\sigma(gc\rightarrow t\rho)\times{\rm BR}(\rho\rightarrow tc)$ can exclude $175~{\rm GeV} \leq \mrho \leq 1000$~GeV for $v_s \lesssim$~1 TeV.

\end{itemize}

In the following we focus on the LHC and study additional signals -- {\em our} smoking guns -- generated by our introduction of DM.
Since we use Ref.~\cite{Bauer:2016rxs}'s Yukawa texture, and since generally $\{ \mrho,~\msigma \} < 2 \mdm$, the branching fractions of the flavon quanta remain the same, and we wish to clarify that adding our DM does not {\em alter} the phenomenology of \cite{Bauer:2016rxs}, only {\em augment} it.

Using the fact that our DM must be detected as missing energy (in association with a visible particle) and that our mediator couples strongest to $b\bar{b}$ and $t\bar{c}$, we will show that searches for heavy quarks + \MET (mono-bottom and mono-top), as well as monojet searches, are our best strategy.
In general, we expect these signals to be difficult to observe.
We may see this from two considerations:
(1) Being $Z_2$-odd, $\chi$ must be pair-produced, either through $\sigma$ production followed by its invisible decay or through an off-shell $\rho$.
The former is kinematically suppressed since $\msigma \simeq v_s$ is heavy, the latter phase-space suppressed,
(2) In most of our signals, the initial state involves the sea quarks $b$ and $c$ that have low PDFs.

It is for the above reasons that we expect to be quite safe from current bounds, as we will show explicitly at the end of this sub-section. 
Due to this lack of constraints, we will restrict ourselves in the following to only a qualitative discussion of our best-case signals.
A more thorough treatment involving careful background estimates and signal-enriching techniques will be dealt in forthcoming work \cite{CAFENR}.
For now, we quote generator-level cross sections at $\sqrt{s}$ = 13 TeV for a point that evades our stringent kaon mixing constraints, $\mdm$ = 760 GeV, $\ychi$ = 1.2, $\mrho$ = 175 GeV, $\msigma$ = 315 GeV.
These cross sections are monojet: 2.2 ab, mono-bottom: 0.4 ab, mono-top: 0.02 ab.

While below we present our LHC signals with DM production through off-shell $\rho$, it must be kept in mind that contributions from processes mediated by a possibly light $\sigma$ may also be relevant.
\\ 

\underline{{\em Monojets}}

\begin{figure}[h!]
\centering 
\includegraphics[scale=0.52]{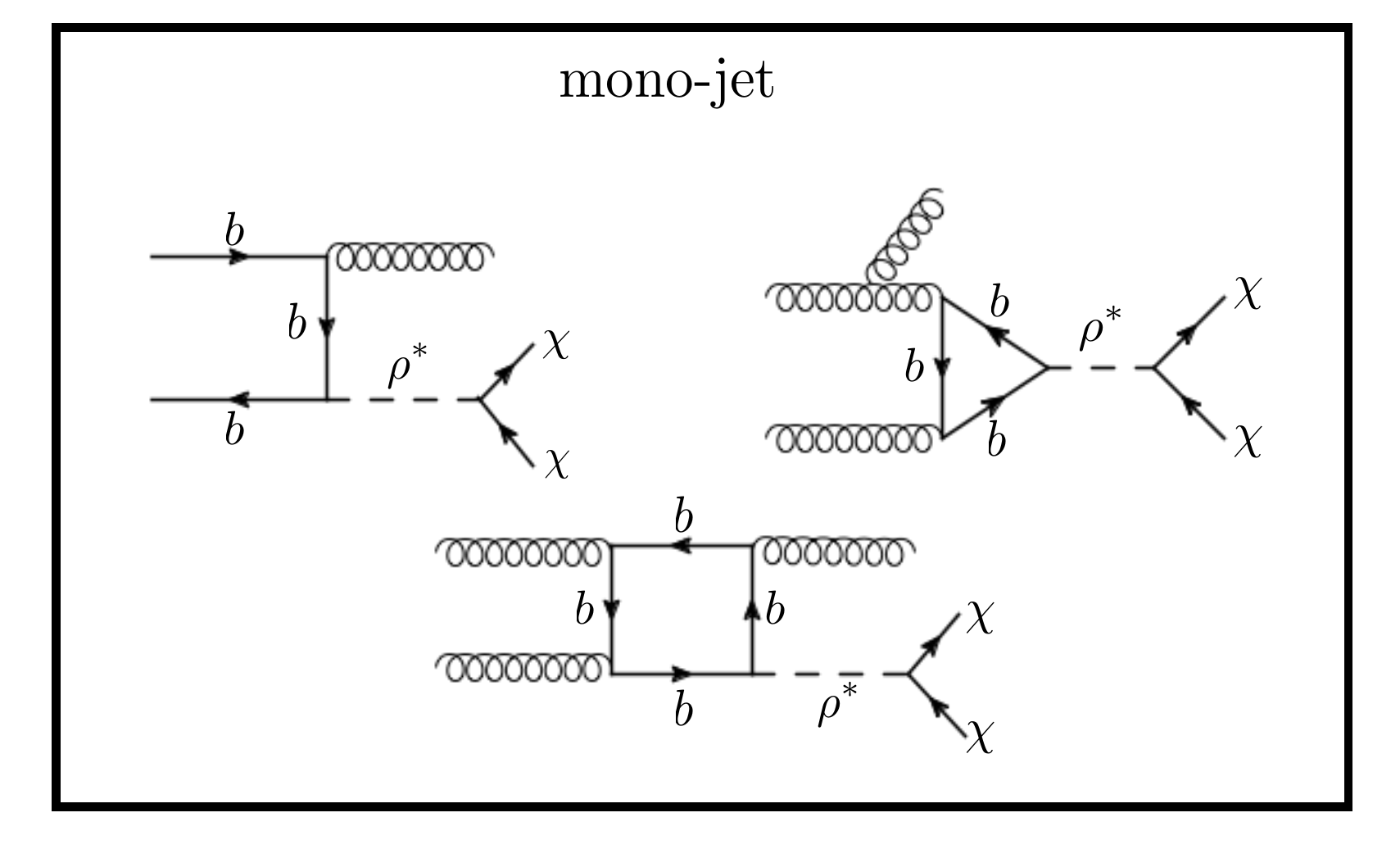}\\
\includegraphics[scale=0.52]{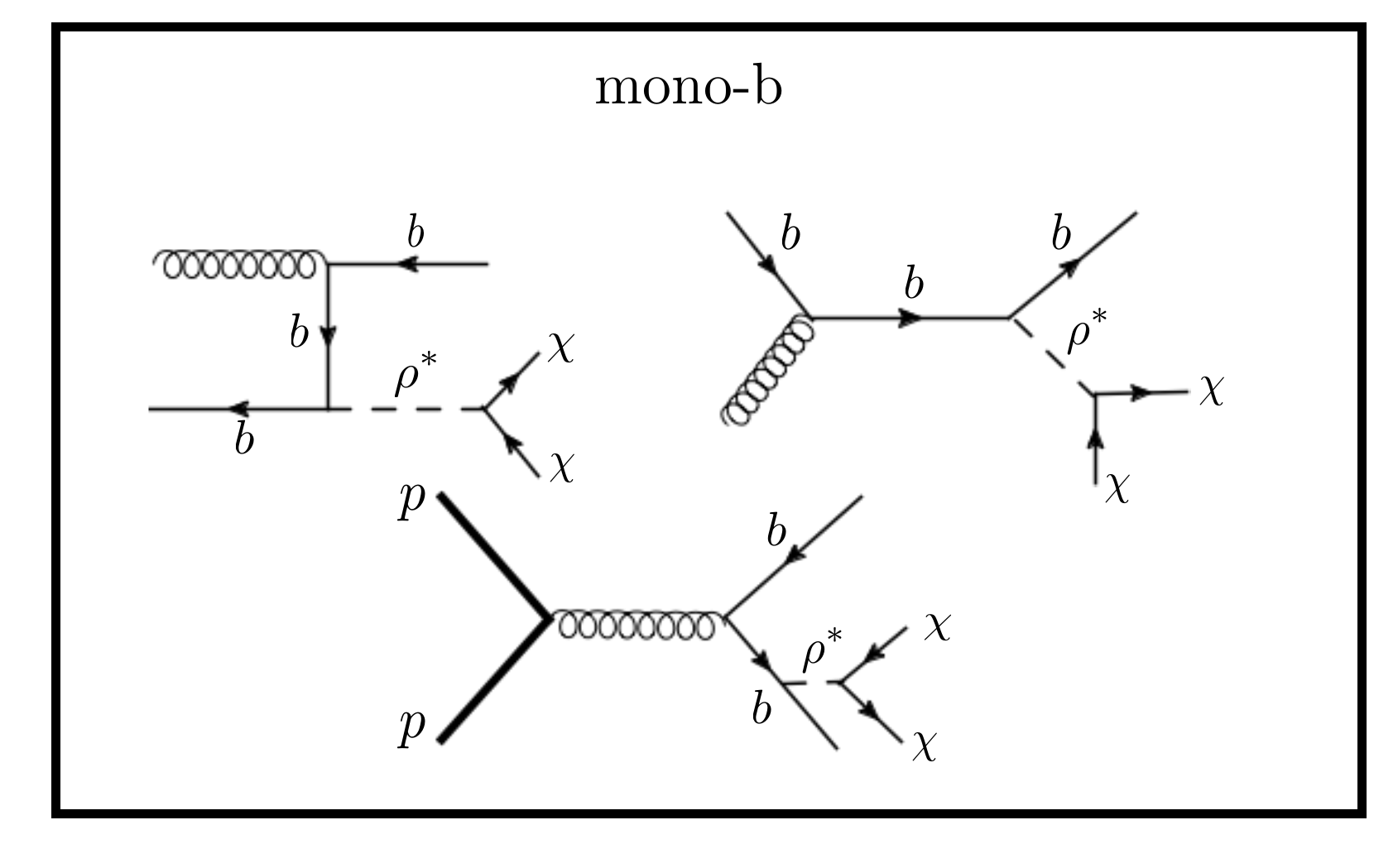}\\
\includegraphics[scale=0.52]{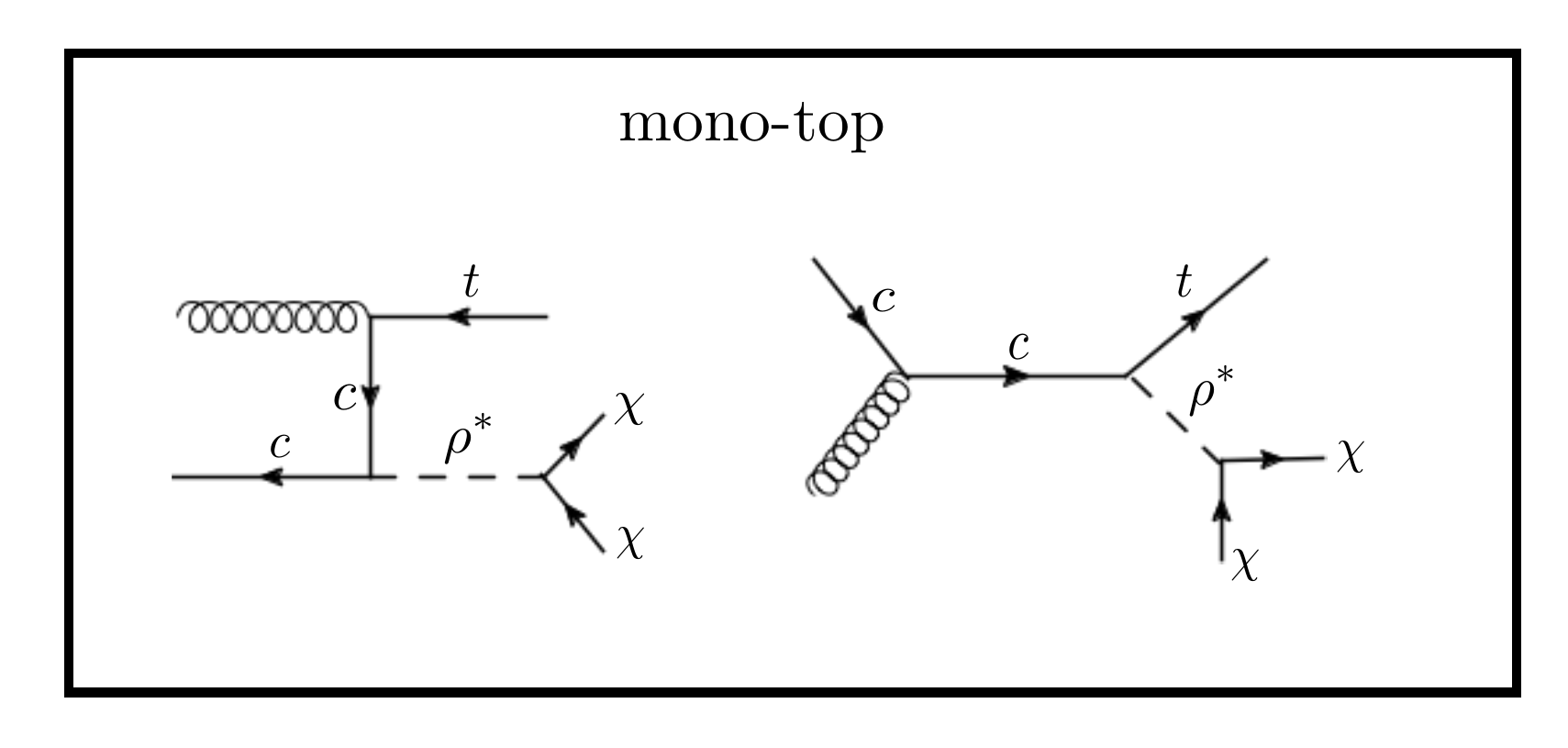}
\caption{Our main signal processes producing mono-jet, mono-$b$ and mono-top signatures.
See text for details on backgrounds and signal-enrichment.
}
\label{fig:monojet}
\end{figure}

Missing transverse energy accompanied by a single jet is a popular channel for LHC DM searches \cite{Aad:2015zva,Aaboud:2016tnv}.
This signal arises in our set-up from the diagrams in Fig.~\ref{fig:monojet}.  
Contributions of the form $q \bar{q} \to \rho^* j$ and those involving lighter quarks in the loop will be negligible because of the weaker couplings.
The main backgrounds are $Z( \nu \bar \nu)+j$, $W( \nu \ell)$ ($\ell$ fakes a jet), $W( \nu \ell)+j$ ($\ell$ is missed).
Another significant but poorly understood background comes from QCD jet mismeasurement, usually minimized by a tight \MET cut.

\underline{{\em Mono-b}}

$b\, g \to \rho^* b$ (Fig.~\ref{fig:monojet}) can contribute to a mono-bottom signal, with subdominant contributions from the flavor-changing processes $q\, g \to \rho^* b$, where $q=d,s$\footnote{ Although light quarks in the initial state will enhance the cross section due to their PDFs, the coupling of $\rho$ to light quarks is so small that these contributions are sub-dominant to $b\, g \to \rho^* b$.};
$b \bar b \rho^*$ production may also contribute when one of the $b$'s is missed. 

The dominant SM backgrounds are still $Z(\nu \bar \nu)/W( \nu \ell)+j/c$, where the ordinary or $c$-tagged jet is misidentified as a $b$ and the $\ell$ is missed.
These contributions, though appearing to be suppressed by mistag rates, outdo direct $b$ production \cite{Lin:2013sca}: $gb \to b Z( \nu \bar \nu)$ is the only irreducible background, and just like the signal process $bg\to \rho^* b$, is PDF suppressed. 
Future improvements in $b$-tagging algorithms may
reduce the $Z/W+j/c$ background, and 
cutting high on \MET would help suppress all the backgrounds.

\underline{\em Mono-top}

The flavor off-diagonal coupling $g_{tc}$ is comparable in size to $g_{bb}$ (see Eq.~(\ref{eq:flavon_coup})) and helps in obtaining a large signal of mono-tops \cite{1106.6199}.
See Fig.~\ref{fig:monojet} for the attendant diagrams.
With the charm PDF higher than the bottom, this channel may be more relevant than mono-$b$. 

Hadronically decaying tops may be particularly advantageous, since the branching ratio is high (=68\%) and since the top mass may be reconstructed from visible final states, aiding in background reduction.
The main background is QCD multijets with mismeasured jets, which can be controlled with a high \MET requirement.  
This may boost the top quark and give near-collinear  final products. 
Hence, instead of three distinct jets, the top may be detected as a single fat jet. 
All this makes our mono-top similar to our mono-jet (in both signal and backgrounds), except here we further demand $m_{\text{fat jet}} \simeq m_t$ and $b$-tagging. 

Having detailed our signals, we now show the safety of our scenario from constraints at the 8 TeV LHC. 
Ref.~\cite{1310.4491} places these bounds for monojet signals on an effective cutoff $\Lambda$ that mediates $\chi$-$q$ interactions ($q = u, d, s, c, b$), which can be recast to our scenario by mapping our parameters to the definition of $\Lambda$.
First, we choose the smallest DM mass spared by kaon mixing: 760 GeV (see the previous two sub-sections).
If $\msigma \sim \mdm$ and $\mrho < \mdm$ (as is true in most regions on our relic surface), the propagator is dominated by the momentum required to pair-produce DM, $2 \mdm$, which is then our cutoff.
Thus 
$\Lambda = 1520$~GeV for both $\sigma$-mediated  
and $\rho$-mediated DM production.
However, Ref.~\cite{1310.4491} finds the tightest bound to only be $\Lambda > 50$~GeV.
Similarly, we can recast the mono-$b$ bound in Ref.~\cite{Lin:2013sca}, which is even weaker: $\Lambda > 90$~GeV.
Finally, mono-top signals studied in Ref.~\cite{1311.6478} set bounds on a model analogous to ours. 
While our $(g_s)_{tc}$ at the parametric point considered above is 0.01,
Ref.~\cite{1311.6478}'s upper limit on the coupling of a pseudoscalar mediator to the up and top quark $\sim O(1)$.
Our actual limit is much weaker, since our $\rho$ and $\sigma$ production must proceed through the sea charm quark in the initial state as opposed to the valence up quark in Ref.~\cite{1311.6478}, and also because our production rates are hurt by a 3-body final state when $\rho$ is in the propagator.  
Analogous to our indirect detection limits, our collider limits are generally weaker than flavor limits because $\mdm \sim v_s$. 
Again similar to indirect detection, we expect the LHC to probe our scenario better at $n > 0$ since $\mdm \propto \epsilon^n v_s$ (Eq.~\ref{eq:genparams}).

Though we have only considered mono-X searches, other signals involving $\rho^* \rightarrow \chi \bar{\chi}$ may be explored.
E.g. a $b \bar{b} +$\MET signal can arise via QCD $b$-pair production with one of the $b$'s radiating a $\rho^* \rightarrow \chi \bar{\chi}$. 
The lack of PDF suppression may bolster the signal, but the background also becomes more significant. 
Moreover, the signal rate falls away for heavy DM. 
These various tensions make this avenue a potentially interesting study.

\section{Conclusions}
\label{sec:concs}

In this paper, we investigated the conditions under which fermionic dark matter embedded in the Froggatt-Nielsen mechanism, with a cutoff scale of $\LFN$, can freeze out to give the observed abundance.
Annihilations to SM species are suppressed by the cutoff scale, while those to the flavon quanta can proceed with $\Oc(1)$ couplings.
If neutral under $\ufn$, DM can undergo $p$-wave-suppressed annihilations to pairs of the CP-even flavon $\sigma$, provided a compressed spectrum is contrived.
If charged under $\ufn$, DM can annihilate freely to a CP-odd flavon $\rho$ + CP-even flavon $\sigma$ in the $s$-wave, as the hierarchy $\mrho \ll \mdm$ can be naturally arranged. 
Flavor constraints on the FN vev $v_s$ and perturbativity of the DM-flavon coupling $\ychi$ together restrict the collective $\ufn$ charge of DM to $\nchi \leq 2$.
Perturbativity also sets upper limits on the FN cutoff $\LFN$, implying that future experiments sensitive to this scale (such as precision low-energy measurements or a 100 TeV collider) may be able to falsify our scenarios. 

We focused on the case of $\nchi = 1/2$, which is viable over a larger parametric region than the cases of higher $\nchi$.
We found that 
while direct detection, indirect detection and LHC searches provide weak constraints,
measurements of the CP-violation parameter $\epsilon_K$ in kaon mixing probe this scenario well.
Future direct detection searches can unearth regions allowed by kaon mixing, but future indirect detection searches cannot.
We also discussed possible DM signatures at LHC Run-2 and strategies for enriching the signal over background.
The clearest signal of our hypothesis would be a triple discovery of the pseudoscalar flavon, CP-even flavon and DM on the ``relic surface" of our parameters, such as along the contours shown in the right panel of Fig.~\ref{fig:ann}.
These possibilities will be explored in greater detail in a forthcoming paper \cite{CAFENR}.
In it we will more closely examine the implications of indirect detection on the entire parameter space in Eq.~\ref{eq:genparams}, studying each $\nchi$ scenario in detail.
We will also undertake a fuller collider study of our LHC signals. 

Our set-up can be trivially extended to include the leptonic FN mechanism, in which case future AMS-02 measurements could become relevant.
Gauging the $\ufn$ symmetry is another possibility, potentially introducing $Z'$ bosons as a DM annihilation channel, and as an avenue for a new set of constraints.
Spin-0 DM takes our analysis into non-trivial directions, since by virtue of the Higgs portal term $|\chi|^2|H|^2$, annihilations could now be shared amongst flavon- and Higgs-pair channels.
We leave all these possibilities for future study.

In summary, it is intriguing that thermal freezeout provides a target for exploring not only the identity of dark matter but also the apparatus behind flavor.

\section*{Acknowledgments}

We are thankful 
to
Fady Bishara
for collaboration in the initial stages of this work and for enlightening Skype sessions, 
to
Antonio Delgado
for spirited discussions and insights into explicit symmetry breaking and global symmetries, 
to 
Wolfgang Altmannshofer,
Martin Bauer,
Chris Kolda,
Adam Martin,
Tilman Plehn,
Oleg Popov,
and
Flip Tanedo
for conversations,
and 
the anonymous referee for helpful comments.
We thank A.~Martin and A.~Delgado for reading the manuscript.
This work was supported by the National
Science Foundation under Grants 
No. PHY-1417118 and No. PHY-1520966.

\appendix  	

\section{Yukawa couplings}
\label{sec:allYukawas}

The quark masses arising from Eq.~\ref{eq:FN1} are
\beq
(m_{u})_{ij} = y_{ij}^{(u)}\epsilon^{\mathcal{Q}_{q_{i}}-\mathcal{Q}_{u_{j}}}\dfrac{v}{\sqrt{2}}~,  
(m_{d})_{ij} = y_{ij}^{(d)}\epsilon^{\mathcal{Q}_{q_{i}}-\mathcal{Q}_{d_{j}}}\dfrac{v}{\sqrt{2}}~, \\
\label{eq:qMass}
\eeq
with the possibility
\begin{align}
\nn \begin{pmatrix}
 \mathcal{Q}_{q_1} & \mathcal{Q}_{q_2} & \mathcal{Q}_{q_3} \\
    \mathcal{Q}_{u} & \mathcal{Q}_{c} & \mathcal{Q}_{t}\\
    \mathcal{Q}_{d} & \mathcal{Q}_{s} & \mathcal{Q}_{b}
  \end{pmatrix}
= 
\begin{pmatrix}
     -3 &  -2 & ~0 \\
    ~~5 & ~~2 & ~0 \\
    ~~4 & ~~3 & ~3
  \end{pmatrix}\
  \end{align}
yielding the correct masses \cite{Bauer:2016rxs}.

The corresponding Yukawa interactions with the physical Higgs are identified as $(Y_{ij}^{u}/\sqrt{2})h\overline{u_{\text{L}i}}u_{\text{R}j}$ and $(Y_{ij}^{d}/\sqrt{2})h\overline{d_{\text{L}i}}d_{\text{R}j}$ with
\begin{equation}
Y_{ij}^{u}\equiv y_{ij}^{(u)}\epsilon^{\mathcal{Q}_{q_{i}}-\mathcal{Q}_{u_{j}}},~~~~~Y_{ij}^{d}\equiv y_{ij}^{(d)}\epsilon^{\mathcal{Q}_{q_{i}}-\mathcal{Q}_{d_{j}}}~.
\label{eq:HiggsYukawas}
\end{equation}

From Eqs.~\ref{eq:vevexpand} and \ref{eq:FN1}, one also obtains the Yukawa interactions 
\beq
 g^q_{ij} \sigma \bar f_{L_i} f_{R_j} + g^q_{ij} \gamma_5 \rho \bar f_{L_i} f_{R_j}
 \label{eq:flavon_coup}
\eeq 
with
\bea
\nn (g_{\sigma}^{u})_{ij} &= i(g_{\rho}^{u})_{ij}=y_{ij}^{u}(\mathcal{Q}_{q_{i}}-\mathcal{Q}_{u_{j}})\epsilon^{(\mathcal{Q}_{q_{i}}-\mathcal{Q}_{u_{j}})}\dfrac{v}{\sqrt{2}v_{s}}, \\
 \nn (g_{\sigma}^{d})_{ij} &= i(g_{\rho}^{d})_{ij}=y_{ij}^{d}(\mathcal{Q}_{q_{i}}-\mathcal{Q}_{d_{j}})\epsilon^{(\mathcal{Q}_{q_{i}}-\mathcal{Q}_{d_{j}})}\dfrac{v}{\sqrt{2}v_{s}}~. \\
\label{eq:YukawasFlavon}
\eea

The $g^{(u,d)}_s$ matrices are brought to the mass basis via the biunitary transformations that diagonalize the Higgs Yukawas $Y_{ij}^{(u,d)}\equiv y_{ij}^{(u,d)}\epsilon^{n_{ij}(m_{ij})}$. Due to the misalignment between the Higgs and flavon Yukawa bases, the $\sigma$ and $\rho$ can mediate flavor-violating interactions that may be subject to meson-mixing constraints.

\section{Dark matter formulae}
\label{app:DMformulae}

This appendix collects formulae used in the calculation of the relic density and direct detection cross sections.
\\

\subsection*{Relic abundance}

The relic abundance is given by \cite{1204.3622}
\begin{equation}
\Omega_{\chi} h^2 = 0.12 \frac{4.4 \times 10^{-26}~\text{cm}^3/\text{s}}{\langle \sigma v \rangle} \left(\frac{x_f}{25}\right) \left(\frac{106}{g_*}\right)^{1/2} ~, 
\label{eq:Relic}
\end{equation}
where $x_f \equiv m_{\chi}/T_{\rm freezeout}$ and $g_{*}$ counts the entropy degrees of freedom at freezeout.
The annihilation cross section $\langle \sigma v \rangle $ is dominated by process $\chi \bar{\chi} \rightarrow \sigma \rho$, and its $s$-wave piece is given by 

\bea
 \nn \langle \sigma v\rangle_{\rm s-wave}  &=& \dfrac{g_{\rho \chi \chi}^{2}}{256\pi m_{\chi}^{4}(4m_{\chi}^{2}-m_{\rho}^{2})^{2}(4m_{\chi}^{2}-m_{\sigma}^{2}-m_{\rho}^{2})^{2}} \\ 
\nn & & \sqrt{ (4m_{\chi}^{2}-m_{\sigma}^{2})^{2}-2m_{\rho}^{2}(4m_{\chi}^{2}+m_{\sigma}^{2})+m_{\rho}^{4} } \\
 \nn &\times& \biggl\{ -g_{\sigma \chi\chi}m_{\rho}^{4}+m_{\rho}^{2}(2m_{\chi}\lambda_{\sigma \rho\rho}+m_{\sigma}^{2}g_{\sigma \chi\chi}) \\
\nn &+& 2m_{\chi}(4m_{\chi}^{2}-m_{\sigma}^{2})(2m_{\chi}g_{\sigma \chi\chi}-\lambda_{\sigma \rho\rho}) \biggr\}^2~, \\
\label{eq:sWave}
\eea
where the couplings $g_{\rho \chi \chi} = g_{\sigma \chi \chi} = y_{\chi}/\sqrt{2}$ and $\lambda_{\sigma \rho\rho} = \msigma^2/v_s$.
As mentioned in Sec.~\ref{subsec:U1DMcharged}, the $p$-wave contribution is found negligible.

\subsection*{Direct detection}

The spin-independent DM-nucleon scattering cross section is given by \cite{1506.03116}

\begin{equation}
 \sigma_{\chi N}^{\text{SI}} = \dfrac{ \mu_{\chi \text{N}}^{2}}{\pi} f_{N}^{2}~,
\label{eq:XSSI}
\end{equation}
with $\mu_{\chi \text{N}}$ the DM-nucleon reduced mass. 
The effective nucleon-DM coupling $f_{N}$ arises from the operators $(\bar{\chi}\chi)(\bar{q}q)$ and $\bar{\chi}\chi  \mathcal{G}^a_{\mu\nu} \mathcal{G}^{a \mu \nu}$ (via heavy quark loops), and is given by \cite{1007.2601}
\begin{equation}
f_{N}= m_N \left(\sum_{q = u,d,s} f_q f^{\rm N}_{T_q} + \sum_{q = c,t,b} \dfrac{2}{27}f_q f_{T_{G}}^{\rm N} c_q \right),
\end{equation}
with the mass fractions in nucleons
\bea
\nn f_{T_{u,d,s}}^{\text{p}}&=&(0.023,0.032,0.020), f_{T_{u,d,s}}^{\text{n}}=(0.017,0.041,0.020),  \\
\nn f_{T_{G}}^{\text{p}}&=&0.925,~~~~~~~\ \ \ \ \ ~~~~~~~f_{T_{G}}^{\text{n}}=0.922~,
\eea
and $c_q$ a QCD correction factor $= 1 + 11 \alpha_s(m_q)/4\pi$ with $(c_c, c_b, c_t) = (1.32, 1.19, 1)$.
The effective quark-DM couplings are given by
\bea
\nn f_q = \left( \dfrac{g_{\sigma \chi \chi}g_{\sigma q q} }{m_q m_{\sigma}^{2}} \right),
\eea
with $g_{\sigma \chi \chi} = \ychi/\sqrt{2}$ and $g_{\sigma q q}$ are the couplings obtained after rotating to the mass basis the couplings given in Eq.~\ref{eq:YukawasFlavon}.


\end{document}